\documentclass{article}

\usepackage{vmargin,amssymb,amsmath,epsfig,latexsym,booktabs,braket,
gensymb,amsbsy,stackrel,slashed}

\newcommand{\na}{n_{A,1}}
\newcommand{\naa}{n_{A,2}}
\newcommand{\nb}{n_{B,3}}
\newcommand{\nc}{n_{C,4}}
\newcommand{\nd}{n_{D,1}}
\newcommand{\nh}{n_{E,b}}
\newcommand{\nv}{n_{E,f}}

\newcommand{\la}{l_A}
\newcommand{\lb}{l_B}
\newcommand{\lc}{l_C}
\newcommand{\ld}{l_D}
\newcommand{\lh}{l_{E,b}}
\newcommand{\lv}{l_{E,f}}
\newcommand{\ep}{\epsilon}
\newcommand{\rar}{\, \rightarrow \,}
\newcommand{\cA}{{\cal A}}
\newcommand{\cN}{{\cal N}}
\newcommand{\cO}{{\cal O}}
\newcommand{\bX}{\bar X}
\newcommand{\Tr}{\mathrm{Tr}}

\newcommand{\beq}{\begin{equation}}
\newcommand{\eeq}{\end{equation}}

\begin{document}

\thispagestyle{empty}
\begin{flushright} HU-EP-22/12
\end{flushright}

\vskip 5 cm

\begin{center}
{\huge \textbf{Tilings and Twist at} $\mathbf{1/N^4}$}

\vskip 1.5 cm

{\large B.~Eden$^\dagger$, T.~Scherdin$^{\dagger \dagger}$}

\vskip 1.25 cm

$^\dagger$ Humboldt-Universit\"at zu Berlin, Institut f\"ur Mathematik \\
$^{\dagger \dagger}$ Humboldt-Universit\"at zu Berlin, Institut f\"ur Physik

\vskip 0.5 cm

E-mail: {\tt eden$@$math.hu-berlin.de, scherdit$@$physik.hu-berlin.de}

\end{center}

\vskip 3 cm

We re-consider operator mixing in the so-called $SU(2)$ sector of $\cN \, = \, 4$ super Yang-Mills theory with gauge group $SU(N)$. Where possible, single-trace operators of moderate length are completed by higher-trace admixtures so as to yield large $N$ tree level eigenstates. 

We are particularly interested in parity pairs with three excitations. Since parity is respected in the mixing, the odd single-trace operators at low length cannot receive too many admixtures. We reproduce the tree-level norms of a set of large $N$ eigenstates up to order $1/N^4$ by integrability methods. This involves evaluating two-point functions on the sphere, the torus, and the double-torus. A perfect match is found as long as descendents are absent from the mixing. 

Using twist to make the descendents appear in the integrability picture immediately leads to the question how to modify the entangled states occurring in the hexagon tessellations. We take a closer look at the double-trace admixtures to the parity even three-excitation operator at length seven, which are both products of a primary state and a descendent. Their two-point functions are sensitive to the twist introduced into the Bethe equations. For transverse scalar excitations we succeed in recovering the corresponding field theory results. For longitudinal magnons our methods fail, pointing at a potential weakness of the formalism.

\newpage

\section{Introduction}

The AdS/CFT correspondence \cite{123} entered a quantitative phase with the \emph{BMN construction} \cite{bmn} which directly linked a class of string theory states to a dual set of composite operators in $\cN \, = \, 4$ super Yang-Mills theory. All elementary fields of this superconformal field theory carry an adjoint representation of a non-Abelian gauge group. The operators in question are products of very many scalar fields of the same type with a few other fields termed \emph{impurities} in between, made gauge invariant and cyclic by a single trace over the gauge group. The one-loop anomalous dimensions of these long operators can be matched with string theory predictions in the same expansion parameter, although this would be small in gauge theory and large for the strings \cite{headrickPlefka}.

In the simplest case, the impurities are all scalars of the same second type. The spectrum of \emph{planar} one-loop anomalous dimensions in such an $SU(2)$ \emph{sector} is equivalent to the energy eigenvalues of the Heisenberg spin chain \cite{Minaza,beiStau}. This observation has opened up an entire research agenda by the name of \emph{integrability} in which one tries to include higher-loop corrections and to extend the integrable system to all types of composite operators.

While the inclusion of higher genus corrections had been discussed in the literature on the BMN operators from the very beginning \cite{headrickPlefka}, it remained an open question for a long time how to address non-planar corrections in the integrability framework \cite{charlotteReview}. The answer came in an unexpected fashion: an efficient scheme for the computation of structure constants was designed in \cite{BKV} and extended to higher-point functions in \cite{cushions,shotaThiago1}. Feynman diagrams for higher-point functions of single-trace operators can be drawn on closed Riemann surfaces with punctures corresponding to the operator insertions. At tree level in configuration space, propagators connecting the operators naturally triangulate the surfaces. Each tile of such a triangulation is given by the \emph{hexagon operator} of \cite{BKV}.

In a systematic large $N$ expansion, single-trace operators will acquire admixtures of higher-trace operators. In \cite{colourDressed} it was pointed that coincidence limits of higher-point functions can be used to describe correlation functions of operators with more than one colour trace. This is one step towards non-planar integrability, and the one that we elaborate on in the present article up to $1/N^4$ corrections. Eventually, it would be desirable to design a formalism capable of calculating the mixing coefficients at non-planar order. This second step will not be addressed here.

We will seek eigenoperators of the conformal group at one loop in the $SU(2)$ sector, albeit without sending the number of elementary fields to infinity. The one-loop problem uniquely fixes the tree level eigenstates as long as the operators are not protected from quantum corrections, i.e. when they have non-vanishing anomalous dimension. We will then study how to combine these operators in a large $N$ expansion. Ideally, we would like to construct
finite $N$ eigenstates; more realistically, we consider the first three terms of a large $N$ expansion in a range of cases.

Building two-point functions of these exact (or $N$ expanded) eigenoperators one has to compute a number of overlaps between their constituents. Here we compare hexagon tessellations to tree level field theory with very good success; problems arise where \emph{descendents} come in, thus operators that are obtained from others by $SU(2)$ raising.

The article is organised as follows: in Section \ref{QFT} we review the construction of field theory eigenoperators using the one-loop \emph{dilatation operator} \cite{stauPlefChristi} and discuss principle features of the spectrum, in particular the appearance of \emph{parity pairs}. In Section \ref{Bethe} we review the integrability formalism for the one-loop spectrum and give a precise map form Bethe wave functions to normalised field theory operators. The set of examples we will explicitly discuss comprises an \emph{exceptional operator} and two \emph{descendents} whose integrability description requires the notion of \emph{twist}, so the modification of the Bethe equations of the Heisenberg chain by a small parameter, see \cite{afsNepo} and references therein. In Section \ref{Hexagons} we introduce the hexagon operator for three-point functions and comment on the effect of twist in this context. Finally, in Section \ref{tessel} we introduce tessellations for higher-point functions and use them to re-calculate all two-point functions related to our set of examples. Section \ref{descCor} is dedicated to two-point functions of double-trace operators one of whose parts is a descendent.

\section{The spectrum in the $SU(2)$ sector from the field theory} \label{QFT}

The $\cN \, = \, 4$ model has the six complex scalar fields $\phi^{[I,J]}, \, I \in \{1,2,3,4\}$ obeying the reality constraint
\beq
\left(\phi^{IJ}\right)^* \, = \, \phi_{IJ} \, = \, \frac{1}{2} \epsilon_{IJKL} \phi^{KL}
\eeq
So three of them, say, $\Phi^i \, = \, \phi^{1,i+1}, \, i \in \{1,2,3\}$ are independent with the reality constraint determining their complex conjugates. To simplify the notation the scalar fields are often called $Z, \, X, \, Y$.

The scalar fields transform in the adjoint representation of a non-Abelian gauge group: $\Phi^i \, = \, \Phi^{ia} \, T^a$ where $T$ is a generator of the local symmetry. For definiteness we assume this \emph{colour group} to be of $SU(N)$ type in this article. We will study large $N$ perturbation theory at tree level w.r.t. to the gauge coupling $g_{YM}$. Gauge invariant composite operators arise as products of the elementary fields under an $SU(N)$ trace. Limiting the scope to $\Phi^1, \, \Phi^2$ or $\Phi^1, \, \Phi^3$ we obtain what is called an $SU(2)$ sector, because the second type of field is obtained from the first by the step operator $\partial_2^3$ or $\partial_2^4$, respectively.

The $\cN \, = \, 4$ model is conjectured to be conformally invariant. Good operators are then the eigenstates of the conformal group, which ought to be orthonormal:
\beq
\langle \cO_i \bar \cO_j \rangle \, = \, 0  \quad : \quad i \, \neq \, j \, , \qquad \qquad \langle \cO_i \bar \cO_i \rangle \, = \, \frac{1}{(x_{ij}^2)^{L + g^2 \, \gamma_{1,i} + \ldots} }\, , \quad g^2 \, = \, \frac{g^2_\mathrm{YM} N}{8 \, \pi^2} \label{orthoState}
\eeq
The exponent in the last formula is the quantum corrected \emph{dimension} of the operator,
of which we have indicated the trivial part given by the length $L$, so the number of elementary fields, and the \emph{one-loop anomalous dimension} $\gamma_1$.

Tree level two-point functions do not provide enough information to fix the eigenstates even at tree level, whereas one-loop renormalisation does. In terms of $\cN \, = \, 1$ supersymmetric Feynman rules this arises from the contraction of the effective vertex
\beq
g^2_\mathrm{YM} \int d^4x \, \Tr( [\bar Z, \bar Y] [Z, Y] ) \label{vertN1}
\eeq
onto the two operators in the two-point functions. 

The concept of the \emph{dilatation operator}\footnote{For a point of view closer to the original BMN literature see also \cite{crook}.} was introduced in \cite{stauPlefChristi}: Wick contraction of the barred fields in the vertex on one of the operators in the two-point function removes a $Z$ and a $Y$, and puts another such pair back, though not in all terms in the same position. This operation defines a linear map, say, $D$ on the spaces of operators with identical length $L$ and number $n$ of $Y$ fields. The one-loop anomalous dimensions $\gamma_i^{(1)}$ are given by the eigenvalues of $D$ because the actual Feynman integral is universal and can be factored out. Note that the $N$ dependence of the eigenvalues can be quite involved whereby it is usually preferable to work perturbatively in $1/N$.

The contraction on the second operator amounts to selecting matrix elements. These do, of course, contain the same information so that one can construct the orthogonal states also from the complete set of tree and one-loop two-point functions, cf. \cite{rome}. Technically, this is harder because a system of quadratic equations on the coefficients in the eigenstates has to be solved.

Now, $SU(N)$ generators satisfy $\Tr(T^a) \, = \, 0$ and the \emph{cutting and sewing rules}
\beq
\Tr(T^c \, A \, T^c \, B) \, = \, \Tr(A) \, \Tr(B) - \frac{1}{N} \Tr(A \, B) \, , \qquad
\Tr(T^c \, A) \, \Tr(T^c \, B) \, = \, \Tr(A \, B) - \frac{1}{N} \Tr(A) \, \Tr(B) \, ,
\eeq
where $A, \, B$ are any products of generators. In particular, contracting the effective vertex onto a \emph{single-trace operator} can cut it up into a \emph{double-trace operator}, while the vertex can do both, sew a \emph{double-trace operator} to a single trace, or to further split it into three traces and so forth. Upon computing with these rules, one finds that contractions of operators with different numbers of traces are suppressed by powers of $1/N$. 

We will try to complete any given leading $N$ single-trace eigenstate $\cO_i^s$ to a series of the form
\beq
\cO_i^s + \left(\frac{a_i^j}{N^2} + \ldots \right) \cO_j^s+ \left(\frac{b_i^k}{N} +  \frac{\hat b_i^k}{N^3} + \ldots \right) \cO_k^d + \left(\frac{c_i^l}{N^2} + \ldots \right) \cO_l^t + \ldots \label{prog}
\eeq
with eigenvalues $\gamma_{1,i}^s + \hat \gamma_{1,i}^s/N^2 + \ldots \, $. All operators in this formula should be leading $N$ eigenstates; with the subscripts $s, \, d, \, t$ we denote the number of traces\footnote{A study of non-planar corrections to anomalous dimensions of operators related to giant gravitons has been initiated in \cite{deMelloKoch}.}. One expects the entire expansion to go in powers of $1/N^2$ with an offset of $1/N$ per additional trace in the admixtures. Last, to avoid trivial rescalings we impose $a_i^i \, = \, 0$ (no sum).

The dilatation operator is a map of the type
\begin{eqnarray}
\cO_i^s & \rar & N \gamma_{1,i}^s \, \cO_i^s + d_i^k \, \cO_k^d \nonumber \\
\cO_k^d & \rar & e_k^j \, \cO_j^s + N \gamma_{1,k}^d \, \cO_k^d + f_k^l \cO_l^t \\
\ldots & & \nonumber
\end{eqnarray}
when all operators are leading N eigenstates. Hence in a hypothetical system with one single- and one double-trace operator we have to solve the eigenvalue equation
\beq
D \left( \cO^s + \frac{b}{N} \cO^d + \ldots \right) \, = \, N \gamma_1^s \, \cO^s + d \, \cO^d + \frac{b}{N} \left( \cO^s + N \gamma_1^d \, \cO^d \right) + \ldots \, = \, (N \, \gamma_1^s + \ldots) \left( \cO^s + \frac{b}{N} \cO^d + \ldots \right)
\eeq
which is trivial at the leading order while we learn from the $O(N^0)$ coefficient of $\cO^d$ that
\beq
b \, = \, \frac{d}{\gamma_1^s - \gamma_1^d} \, .
\eeq
Consequently, if there is degeneracy $\gamma_1^s \, = \, \gamma_1^d$ in the system while $d \, \neq \, 0$, we will not succeed in constructing the desired completion \eqref{prog} of the single-trace eigenstate\footnote{In this situation further problems arise, e.g. the expansion in terms of the coupling constant does not commute over that in terms of the genus counting parameter \cite{stauPlefChristi}. Our point of view is to expand in the coupling first.} \cite{headrickPlefka,moreDegenerate,stauPlefChristi}. In fact, the whole program seems doomed --- wouldn't a similar problem arise, if anywhere in the expansion the one-loop anomalous dimension of any two multiple-trace operators were degenerate? Despite the fact that our simple experiment with only one operator of each type quickly becomes inconsistent going to higher orders, it does indicate that only the one-loop anomalous dimension of any candidate admixture and that of the single-trace operator to be completed have to be distinct. This condition is not impossible to satisfy because the anomalous dimensions of the higher states are roots of fairly complicated characteristic polynomials and are therefore not too likely to reproduce previously encountered values. In fact, scanning the spectrum of single-trace operators up to length ten and four excitations to the order indicated in \eqref{prog} we observed a breakdown of the ansatz only for cases with $\gamma_1^s \, \in \{4, 5, 6\}$. 

Because there are as many eigenvectors in the single-trace sector of given $L, \, n$ as there are states, we can always re-write the admixtures as products of leading $N$ single-trace eigenstates. Incidentally, to leading $N$ the one-loop anomalous dimension of such a product is the sum of that of its factors.

The one-loop mixing problem has the following features:
\begin{itemize}
\item States of different one-loop anomalous dimension will be orthogonal at tree level.
\item We can act on any state with length $L$ and $n$ excitations with the  step operator $\partial_2^4$ by global differentiation to produce a \emph{descendent} of the same length but with one more excitation. If non-vanishing, the descendent will be an eigenvector and has the same anomalous dimension as the original operator. States that are not descendents are called \emph{primary}.
\item All the admixtures to renormalised single-trace operators (i.e. with $\gamma_1^s \, \neq \, 0$) also have non-vanishing anomalous dimensions. Since the one-loop anomalous dimensions of all operators are non-negative, at least one factor in each multi-trace operator is renormalised. This property should be rooted in supersymmetry because non-renormalised operators are members of shorter supermultiplets that could not consistently be added in. 
\item At higher length and excitation number the leading $N$ single-trace spectrum contains \emph{degenerate pairs}. As a basis of such a $2 \times 2$ cell we can choose the even and odd part under \emph{parity}, i.e. reversal of the trace as in $\Tr(ZZYZYY) \rar \Tr(ZZYYZY)$. Parity is a strict rule \cite{stauPlefChristi,paulDHoker}: the admixtures all have the same parity as the single-trace state we complete.
\item When the two states of definite parity in a leading $N$ degenerate pair can be completed by admixtures as advocated in \eqref{prog} the degeneracy will be lifted by the higher $N$ corrections\footnote{See also \cite{chZABJM} for similar results in other instances of the AdS/CFT correspondence.} \cite{stauPlefChristi}.
\end{itemize}
All states with $L \, = \, 2,3$ are \emph{non-renormalised} or \emph{protected}, i.e. have $\gamma_1^s \, = \, 0$. At every $(L,n)$ there will be one protected single-trace state; at the multi-trace level there are in general several protected operators according to how $L$ is partitioned into shorter parts and how the excitations are distributed over these. Especially at short length most states are parity even. Therefore the mixing in the odd sector is significantly simpler.

In fact, all states with $n \, = \, 0,1$ $Y$ fields are protected. Also, all states with $n \, = \, 2$ are parity even. We are therefore particularly interested in the next simplest case $n \, = \, 3$. Scanning states of length up to and including $L \, = \, 9$ we are able to point out all properties that are new in the integrability picture w.r.t. to the previously studied operators with two $Y$ fields \cite{colourDressed}. 

As an illustration, let us comment on the states of length 5: in the protected sector we have the single-trace operators
\beq
\frac{1}{\sqrt{5}} \Tr(Z^5), \, \Tr(Z^4 Y), \, \frac{1}{\sqrt{2}} \Tr(ZZ\{Z,Y\}Y), \, \frac{1}{\sqrt{2}} \Tr(Z\{Z,Y\}YY), \, \Tr(Z Y^4), \, \frac{1}{5} \Tr(Y^5) \, .
\eeq
All of these can be generated by derivatives on the first operator in the list. Hence they are \emph{vacuum descendents}. Then there are two states of one-loop anomalous dimension $\gamma_1^s \, = \, 4$:
\beq
\frac{1}{\sqrt{2}} Tr(ZZ[Z,Y]Y), \, \frac{1}{\sqrt{2}} \Tr(Z[Z,Y]YY) \, ,
\eeq
which we will refer to as (5,2) and (5,3), respectively. The (5,3) operator is a descendent of the (5,2) case. Double-trace states must be built from factors of length 2 and 3. They are all protected at leading order in $N$ since their constituents are. We can conclude that both renormalised states, (5,2) and (5,3), respectively, receive no admixtures. As a consequence, their one-loop anomalous dimension will receive no corrections in $1/N$. Notice that all length 5 states are parity even.

Beyond (5,2) and (5,3) we will encounter the following operators in our study:
\begin{enumerate}
\item The first parity odd operator, which comes at $(6,3)$:
\beq
(6,3)^e \, = \, \frac{i}{\sqrt{2}} \Tr(ZZY[Z,Y]Y) \label{op63e}
\eeq
For want of other parity odd states at this length this is a true eigenstate of the mixing problem also at finite $N$, its anomalous dimension $\gamma_1^s \, = \, 6$  does not receive corrections in $1/N$. We have given the operator the label $e$ for \emph{exceptional} for reasons that will become apparent in the next section.
\item At length 7, a first degenerate pair of primaries with $\gamma_1^s \, = \, 5$ appears with three $Y$ fields:
\begin{eqnarray}
(7,3)^- & = & \frac{i}{\sqrt{2}} \, (0,-1,1,0,0) \, , \label{op73m} \\
(7,3)^+ & = & \frac{\sqrt{2}}{\sqrt{15}} \left(-1,\frac{3}{2},\frac{3}{2},-1,-1\right) \label{op73p}
\end{eqnarray}
The coefficient vectors in these formulae refer to the basis
\beq
\Tr(Z^4 Y^3), \, \Tr(ZZZYZYY), \, \Tr(ZZZYYZY), \, \Tr(ZZYZZYY), \, \Tr(ZZYZYZY) \, . \label{bas73}
\eeq
A remark on the norms is in order: we take out a number coefficient so that the leading $N$ part of the two-point functions will be $N^L$. This will be useful in comparing to integrability.

The parity odd case above experiences no mixing, while the parity even one could have the double-trace admixtures $(5,3)(2,0), \, (5,2)(2,1), \, (4,2)(3,1)$. The third of these is a descendent. It does in fact not enter the mixing. The other two come in the special combination
\beq
\cO^\perp \, = \, \frac{1}{\sqrt{3}} \left( \sqrt{2} \, (5,3)(2,0) - (5,2)(2,1) \right) \, .
\eeq
We remark that $\cO^\perp$ is a primary state: it is exactly tree orthogonal to the descendent $\partial_2^4 \, (5,2)(2,0)$. Thus the mixing takes place exclusively between the primaries at $(7,3)$. For now another property is more important: $(7,3)^+$ and $\cO^\perp$ map into each other under the dilatation operation $D$:
\beq
\frac{D}{N} \left( \begin{array}{c} (7,3)^+ \\[1 mm] \cO^\perp \end{array} \right) \, = \, \left( \begin{array}{cc} 5 & -\frac{2 \sqrt{5}}{N} \\[1 mm] -\frac{4 \sqrt{5}}{N} & 4 \end{array} \right) \left( \begin{array}{c} (7,3)^+ \\[1 mm] \cO^\perp \end{array} \right) \label{nonHerm}
\eeq
Notice that this matrix is not symmetric, whereby its left- and right-eigenvectors are distinct. Worse, the members of either set are not tree orthogonal in contradiction to the definition \eqref{orthoState}. An approriate change of basis is discussed in Appendix A.

Here we content ourselves with constructing an eigenoperator $(73)^+ + \, b \, \cO^\perp$ determining the left-eigenvectors $(1,b)$ of the matrix in the last equation:
\beq
\gamma_1 \, = \, \frac{1}{2} \left(9 \pm \sqrt{1 + \frac{160}{N^2}}\right) \, , \qquad b \, = \, \frac{N}{8 \, \sqrt{5}} \left( 1 \mp \sqrt{1 + \frac{160}{N^2}} \right) \label{eigen73}
\eeq
The upper sign in front of the root corresponds to the mixing problem in \eqref{prog} with
\beq
\gamma_1^s \, = \, 5, \, \quad \hat \gamma_1^s \, = \, 40, \qquad b_1 \, = \, -2 \, \sqrt{5}, \quad \hat b_1 \, = \, 80 \, \sqrt{5}
\eeq
while the other sign yields an eigenoperator in which the double-trace part is dominant. Notice that the factor $\sqrt{5}$ arises from the unit normalisation of the operators. To convey the central point of this discussion we juxtapose the anomalous dimensions of the ($N$ corrected) single-trace operators:
\beq
(7,3)^- \, : \, \gamma_1 \, = \, 5 \, , \qquad (7,3)^+ \, : \, \gamma_1 \, = \, 5 + \frac{40}{N^2} + \ldots
\eeq
Hence the $N$ dependence lifts the degeneracy of the eigenvalues, and the parity eigenstates are the good operators at finite $N$ or in a large $N$ expansion. These results were originally derived in \cite{ryzhov,stauPlefChristi} and recently recovered in \cite{whoBeta}.
\item At length 8, there are three primaries with three $Y$'s:
\begin{eqnarray}
\gamma_1^s \, = \, 6 \quad : \quad (8,3)^e \, & = & \frac{i}{2} \left( \Tr(Z^4 Y [Z,Y],Y) - \Tr(Z Z [Z,Y] Z Z Y Y) \right) \, , \nonumber \\
\gamma_1^s \, = \, 4 \quad : \quad (8,3)^+ & = & \frac{1}{2} \left( \Tr(Z^5 Y^3) - \Tr(Z^4 Y \{Z,Y\} Y) + \Tr(ZZYZZYZY) \right) \, \\
\gamma_1^s \, = \, 4 \quad : \quad (8,3)^- & = & \frac{i}{2} \left(  \Tr(Z^4 Y [Z,Y],Y) + \Tr(Z Z [Z,Y] Z Z Y Y) \right) \, . \nonumber
\end{eqnarray}
The leading $N$ degeneracy of $\gamma_1^s \, = \, 6, 4$ with that of $(6,3)^e (2,0)$ and $(5,3)(3,1), \, (5,2)(2,1)$, respectively, prevents us from discussing the first two cases in this list. On the other hand, $(8,3)^-$ is an exact eigenoperator of $D$; it could mix with $(6,3)^e (2,0)$ but does not. In Appendix A we further comment on this example, too.
\item At $L \, = \, 9, \, n \, = \, 3$ the primary single-trace operators come in three degenerate pairs. The three parity even states have a host of admixtures including all four possible triple-trace operators carrying leading $N$ anomalous dimension. Interestingly, there is re-mixing also amongst the parity even primaries at the level of the $a_2/N^2$ coefficients whereas single-trace descendent states do not come in. 

The parity odd single-trace states $(9,3)_i^-$ are
\begin{eqnarray}
\gamma_{1,1}^s \, = \, 8.25342 \, , \qquad (9,3)_1^- & = & i \, ( 0.0532393, -0.17321, 0.683494) \, , \nonumber \\
\gamma_{1,2}^s \, = \, 5.51997 \, , \qquad (9,3)_2^- & = & i \, ( 0.616784, -0.320709, -0.129317) \, , \label{odd93} \\
\gamma_{1,3}^s \, = \, 3.22661 \, , \qquad (9,3)_3^- & = & i \, ( 0.341676,  0.605924,  0.126938 ) \, , \nonumber
\end{eqnarray}
where the coefficient vectors refer to the three parity odd combinations
\beq
\Tr(Z^5 Y [Z,Y]Y), \, \Tr( Z^4 Y [Z^2, Y] Y), \, \Tr(Z^3Y Z [Z,Y] Z Y) \, .
\eeq
Curiously the one-loop anomalous dimensions of the descendent states $6.82843, 4, 1.17157, 0$ lie in between.

The $(9,3)_i^-$ operators can and do mix with $(6,3)^e (3,0), \, (7,3)^- (2,0)$ whereas there are no parity-odd triple-trace operators at length 9. Like their even cousins they display re-mixing among themselves at $O(N^{-2})$. Up to $1/N^3$ we obtain
\begin{eqnarray}
&& (9,3)_1^- + \frac{0.303442}{N^2} (9,3)_2^- - \frac{0.0353558}{N^2} (9,3)_3^-  \nonumber \\
&& \quad + \left(\frac{0.738458}{N} - \frac{1.72311}{N^3} \right) (6,3)^e (3,0) + \left( \frac{0.461192}{N} - \frac{1.32898}{N^3}\right) (7,3)^- (2,0) + \ldots \, , \nonumber \\
&& (9,3)_2^- + \frac{49.9031}{N^2} (9,3)_1^- + \frac{5.98334}{N^2} (9,3)_3^-  \label{com93} \\
&& \quad - \left(\frac{14.3515}{N} - \frac{1041.83}{N^3} \right) (6,3)^e (3,0) -\left( \frac{4.00861}{N} + \frac{189.762}{N^3}\right) (7,3)^- (2,0) + \ldots \, , \nonumber \\
&& (9,3)_3^- - \frac{2.33609}{N^2} (9,3)_1^- + \frac{1.47962}{N^2} (9,3)_2^-  \nonumber \\
&& \quad + \left(\frac{0.700161}{N} - \frac{1.85918}{N^3} \right) (6,3)^e (3,0) + \left( \frac{1.31087}{N} + \frac{4.92937}{N^3}\right) (7,3)^- (2,0) + \ldots \, , \nonumber
\end{eqnarray}
\end{enumerate}
with the anomalous dimensions
\begin{eqnarray}
\gamma_{1,1}^s \, = \, 8.25342 + \frac{8.18190}{N^2} + \ldots \, , \nonumber \\
\gamma_{1,2}^s \, = \, 5.51997 - \frac{39.8239}{N^2} + \ldots \, ,\\
\gamma_{1,3}^s \, = \, 3.22661 + \frac{1.64201}{N^2} + \ldots  \, .  \nonumber
\end{eqnarray}
For comparison, the parity even partners\footnote{To the same accuracy, all six single-trace primaries are given in Appendix A. We do not display the admixtures.} acquire the one-loop anomalous dimensions
\begin{eqnarray}
\gamma_{1,1}^s \, = \, 8.25342 + \frac{53.8338}{N^2} + \ldots \, , \nonumber \\
\gamma_{1,2}^s \, = \, 5.51997 + \frac{72.4183}{N^2} + \ldots \, , \\
\gamma_{1,3}^s \, = \, 3.22661 + \frac{50.2636}{N^2} + \ldots  \, .  \nonumber
\end{eqnarray}
In the main part of the article we will look at two-point functions of these operators up to $1/N^4$ by integrability methods, that is via hexagon tessellations \cite{BKV,cushions,shotaThiago1}. It was advocated in \cite{colourDressed} that each $m$-trace operator can be realised as a coincidence limit of $m$ vertices of a triangulation of a Riemann surface whose genus matches the order in $N$, i.e. a sphere at leading order, a torus for $1/N^2$ corrections etc. One purpose of the present work is to check this conjecture on a wider and more complicated range of examples. Given the list of cases above we will have to analyse
\begin{itemize}
\item single-trace/single-trace: two-point functions on the sphere, the torus, and the double torus
\item single-trace/double-trace: three-point functions in a point identification limit on the sphere and on the torus
\item double-trace/double-trace: four-point functions in a double coincidence limit on the sphere; the relevant torus part can be read out of previous work \cite{colourDressed}.
\end{itemize}
The focus is not so much on the actual coefficients of the first few terms in the large $N$ expansion of the two-point functions; like the mixing coefficients for the $(9,3)_i^-$ operators these are some rather uninspiring numbers, cf. Appendix A. Instead we embark on checking the hexagon amplitudes against the field theory results for every colour factor, so whether the integrability construction can reproduce field theory term by term.

\section{The spectrum problem from integrability} \label{Bethe}

The leading $N$ part of $D$ arises by nearest neighbour interactions, so when the vertex \eqref{vertN1} is contracted onto two adjacent fields within the same colour trace. In this case the action is proportional to $\mathbb{I} - \mathbb{P}$ (i.e. identity minus permutation), which is the Heisenberg $XXX$ Hamiltonian acting on a chain of down spins $Z$ and up spins $Y$ \cite{Minaza,beiStau}. The leading $N$ eigenstates of our map $D$ in the single-trace sector are therefore the energy eigenstates of the Heisenberg spin chain. The famous Bethe ansatz can be used to construct these: the up spins or $Y$ fields are viewed as excitations moving along the sites of a chain of length $L$. Every such \emph{magnon} is given a momentum $p$ with which it moves along the chain. The step operator
\beq
e^{i \, p} \, = \, \frac{u+\frac{i}{2}}{u - \frac{i}{2}} \, , \qquad u \, = \, \frac{1}{2}\cot\left(\frac{p}{2}\right)
\eeq
takes it from one site of the chain to the next. It is customary to use the \emph{rapidity} (or \emph{Bethe root}) $u$ instead of the momentum $p$ in order to obtain algebraic equations. Pulling a magnon over another one induces a phase
\beq
S_{jk} \, = \, \frac{u_j - u_k + i}{u_j - u_k - i} \label{SU2mat}
\eeq
which goes by the name of \emph{scattering matrix}. The Bethe-Yang equations
\beq
e^{i \, p_j \, L} \prod_{k \neq j} S_{jk} \, = \, 1 \, , \qquad j \in \{1 \ldots n \} \label{bethe}
\eeq
express that transporting a magnon once around the chain should not have any effect because the original configuration is recovered. Translation invariance around the chain implies the usual momentum conservation constraint $\sum_j p_j \, = \, 0$ or in terms of the Bethe rapidities
\beq
\prod_{j=1}^n \frac{u_j+\frac{i}{2}}{u_j - \frac{i}{2}} \, = \, 1 \, .
\eeq
Finally, the energy is given by
\beq
\gamma_1^s \, = \, \sum_{j=1}^n \frac{1}{u_j^2 + \frac{1}{4}} \, .
\eeq
An all $Z$ state --- so one without any magnon --- has energy 0. For one magnon the situation is not much different: the zero momentum constraint implies $p_1 \, = \, 0$ and so $u_1$ is infinite such that again $\gamma_1^s \, = \, 0$. For $n \, > \, \lfloor L/2 \rfloor$ excitations we should discover the same spectrum as for $L-n$ excitations because we might then regard $Y$ as a vacuum site and $Z$ as an excitation. These arguments rule out states of non-vanishing energy for $L \, = \, 2,3$.

Normally, states with non-vanishing energy are given by a set of distinct, finite rapidities.
The first non-trivial case we encounter is (4,2) with Bethe rapidities $\pm 1/\sqrt{12}$ and energy 6. According to the above all other length 4 states are protected.

Let us have a look at the length 5 problem: in the field theory the two states with anomalous dimension 4 are (5,2) and (5,3) as defined in the last section. For the (5,2) state we have the regular non-degenerate solution $u_1 \, = \, 1/2 \, = \, -u_2$ with energy 4. For the (5,3) case there is no regular solution. Yet, one can put, say, $u_3 \, = \, \infty$ whereby it scales out of the equations reducing them to the (5,2) case. One thus has the three rapidities $\pm 1/2, \, \infty$ similar to what we said about the one-magnon case at any length. Indeed, the one-magnon case at length $L$ is always a descendent of the spin chain vacuum $\Tr(Z^L)$. In both descendent cases, $(L,1)$ and (5,3) we found an infinite Bethe root.

Descendents states can be systematically obtained introducing \emph{twist} into the Bethe equations, see \cite{afsNepo} and the references given there. In the present work we do so by an extra factor $e^{i \, \ep}$ in every Bethe equation:
\beq
e^{i \, \ep} e^{i \, p_j \, L} \prod_{k \neq j} S_{jk} \, = \, 1 \label{twistedBethe}
\eeq
The coupled equations are then solved order by order in a series expansion in $\ep$. The two states we need for the set of correlators chosen in the last section are $L \, = \, 2, \, n \, = \, 1$ with a single rapidity
\beq
u_1 \, = \, -\frac{2}{\ep} + \frac{1}{24} \ep + \frac{1}{5760} \ep^3 + \ldots
\eeq
and $L\, = \, 5, \, n \, = \, 3$:
\begin{eqnarray}
u_1 & = & -\frac{1}{2} - \frac{1}{10} \ep - \frac{11}{80} \ep^2- \frac{761}{6000} \ep^3 - \frac{12829}{96000} \ep^4 - \frac{453889}{3000000} \ep^5 + \ldots \, , \nonumber \\ 
u_2 & = &\phantom{-}\frac{1}{2} - \frac{1}{10} \ep + \frac{11}{80} \ep^2 - \frac{761}{6000} \ep^3 + \frac{12829}{96000} \ep^4 - \frac{453889}{3000000} \ep^5 + \ldots \, , \\ 
u_3 & = &  -\frac{1}{\ep} + \frac{29}{60} \ep + \frac{4657}{18000} \ep^3 + \ldots \, . \nonumber
\end{eqnarray}
In the $\ep \rar 0$ limit we do indeed obtain the set of rapidities of the primary (e.g. (20) and (52), respectively) and an infinite rapidity. 

Beyond missing the descendents, the Bethe equations \eqref{bethe} do not have regular solutions corresponding to certain exceptional operators \cite{beiStau} like $(6,3)^e, \, (8,3)^e$ etc. Introducing twist we can construct such solutions, too. For instance, for the $(6,3)^e$ operator we have
\begin{eqnarray}
u_1 & = & -\frac{i}{2} + \frac{1}{3} \ep - \frac{19}{432} \ep^3 + \frac{811}{34560} \ep^5 - \frac{i}
   {243} \ep^6 - \frac{262357}{17418240} \ep^7 + \frac{11 \, i}{1944} \ep^8 + \ldots \, , \nonumber \\
u_2 & = &  \phantom{-} \frac{i}{2} + \frac{1}{3} \ep - \frac{19}{432} \ep^3 + \frac{811}{34560} \ep^5 + \frac{i}
   {243} \ep^6 - \frac{262357}{17418240} \ep^7 - \frac{11 \, i}{1944} \ep^8 + \ldots \, , \\
u_3 & = & -\frac{1}{24} \ep + \frac{97}{3456} \ep^3 - \frac{2843}{138240} \ep^5 + \frac{4889627}{278691840} \ep^7 + \ldots \nonumber
\end{eqnarray}
Looking at the leading terms $\pm i/2, \, 0$ we see that the first two are simple poles/zeroes of the momentum factor which is raised to the sixth power in the Bethe equations, whereas the $u_1 - u_2 + i \, = \, O(\ep^6)$ so that $S_{12}$ can offset this singularity. The energy of the solution is $6 + O(\ep^2)$ as expected for $(6,3)^e$. There are similar solutions at any even length.

For $n \, = \, 2$ the zero momentum constraint always implies $u_2 \, = \, - u_1$. Degenerate pairs arise for sets of roots that do not go into themselves under $\{ u_i \} \rar \{- u_i\}$, which
can occur only if $n \, \geq \, 3$. The (undeformed) Bethe equations, the momentum constraint, and the energy are invariant under the simultaneous sign flip of all rapidities so that asymmetric root sets do come in pairs. Yet, the two sets of rapidities do yield two distinct states. 

Let $0 \, < \, m_j \, \leq \, L$ denote the position of the magnon $j$ with rapidity $u_j$ along the chain. The \emph{Bethe wave function} is
\beq
\psi(L,n) \, = \, \sum_{m_1, \ldots, m_n \, = \, 1}^L \ \prod_{j \, = \, 1}^n e^{i \, p_j \, m_j} \ \prod_{j < k} T_{jk} \ \, |m_1 \ldots m_j\rangle \, , \qquad \qquad T_{jk} \, = \, \Biggl\{ \begin{array}{ccl} m_j \, < \, m_k & : & 1 \\ m_j \, = \, m_k & : & 0 \\ m_j \, > \, m_k & : & S_{jk} \end{array}
\eeq
where the ket state has $Y$ magnons at the sites $m_j$ and otherwise $Z$'s. In this definition the chain is a priori open with a designated beginning and end. The magnons do not yet have definite rapidities. The Bethe equations assure that we can close the chain and so make it cyclic. To that end the rapidities must solve \eqref{bethe}.

A normalised, cyclic state is given by \cite{Korepin}
\beq
{\cal O} \, = \, \frac{\psi(L,n)}{\sqrt{ {\cal G} \, L \, \prod_{j \, < \, k} S_{jk} \, \prod_j (u_j^2 + \frac{1}{4})}} \, . \label{normWave}
\eeq
Obviously, any permutation of the rapidities of a Bethe solution is also a solution. The normalised state does not depend on the ordering of the rapidities as long as the same one is chosen in the wave function as well as the phase factor (the product of $S$ matrices) in the denominator. The last missing piece of information is the definition of the \emph{Gaudin norm}:
\beq
{\cal G} \, = \, \mathrm{Det} \ \phi_{jk} \, , \qquad \phi_{jk} \, = \, \frac{\partial  \log \left( e^{i \, p_j \, L } \prod_{l \, \neq \, j} S_{jl} \right)}{\partial \, u_k}
\eeq
All of this perfectly works for the states involving twist, to leading order in $\ep$. By way of example, in our simplest case the state is
\beq
(2,1) \, = \, \frac{\Tr(YZ) + \Tr(ZY)}{\sqrt{ \frac{\ep^2}{2} * 2 * 1 *  \frac{4}{\ep^2} }} \, = \, \Tr(ZY)
\eeq
as expected for the unit norm descendent in field theory. For the length 5 descendent we find
\beq
(5,3) \, = \, \frac{ 5(1+i) ( \Tr(ZZYYY) - \Tr(ZYZYY) ) }{\sqrt{ (80 \, \ep^2) * 5 * i * \frac{1}{4 \, \ep^2}}} \, = \, \frac{1}{\sqrt{2}} \Tr(Z[Z,Y]YY)
\eeq
and finally
\beq
(6,3)^e \, = \, \frac{\frac{486 \, i}{\ep^5} (\Tr(ZZYZYY) - \Tr(ZZYYZY))}{ \sqrt{ \frac{11664}{\ep^6} * 6 * \frac{243}{\ep^6} * \frac{\ep^2}{36} }} \, = \, \frac{i}{\sqrt{2}} \Tr(ZZY[Z,Y]Y) \, . \label{op63eInt}
\eeq
For the degenerate pairs, the situation is particularly interesting: let us consider the (7,3) states normalised as in \eqref{normWave}. With $j \, = \, 1,2$ we find
\beq
(7,3)^j = \frac{1}{\sqrt{15}} \left(-1,\frac{3}{2},\frac{3}{2}, -1, -1\right) - (-1)^j \frac{i}{2} \left( 0, -1, 1, 0, 0 \right) \label{pair7}
\eeq
in the basis \eqref{bas73} so that comparing to \eqref{op73m}, \eqref{op73p}
\beq
(7,3)^\pm \, = \, \frac{1}{\sqrt{2}} \bigl( (7,3)^1 \pm (7,3)^2 \bigr) \, .
\eeq
Similarly, $(8,3)^\pm \, = \, \bigl( (8,3)^1 \pm (8,3)^2 \bigr)/\sqrt{2}$. In fact, a relation
\beq
\cO^\pm \, \propto \, \frac{1}{\sqrt{2}} (\cO^1 \pm \cO^2)
\eeq
seems to hold in general, where $\cO^\pm$ have purely real/imaginary coefficients and are othonormal at leading $N$.

Last, let us comment on the degeneracy of the spectrum in the $SU(2)$ sector. The two-magnon situation is particularly prone to falling victim to this problem: as the zero momentum constraint implies $u_2 \, = \, - u_1$ the scattering matrix $S_{12}$ becomes equal to $e^{-i \, p}$. The one remaining Bethe equation is 
\beq
e^{i \, p \, (L-1)} \, = \, 1 \qquad \Rightarrow \qquad p \, = \, 2 \, \pi \frac{m}{L-1}, \, m \, \in \{0 \ldots \lfloor(L-1)/2\rfloor \} \, .
\eeq
As a consequence, there are degenerate momenta between the $L$ and the $L'$ problem with two magnons whenever $L-1, \, L'-1$ are not co-prime. For instance, the renormalised (42) state has $L-1 \, = 3, \, m \, = \, 1$ and there is a length 7 two magnon state with $L'-1 \, = \, 6, \, m \, = 2$ with the same energy $\gamma_1^s \, = \, 6$. At length 8 the operators
$(4,2)(4,0)$ and $(4,2)(2,0)(2,0)$ have degenerate leading $N$ first anomalous dimension, at length 9 even $(4,2)(5,0), \, (4,2)(3,0)(2,0), \, (7,2)(2,0)$, and all of these operators are parity even so that they can mix. These are not even the only degeneracies in the $SU(2)$ sector spectrum --- e.g. $(6,3)^e$ has $\gamma_1^s \, = \, 6$ exactly as the primary (4,2) state. This made us worry about the success of our program of completing single-trace eigenstates by higher-trace admixtures. Fortunately, the spectrum of higher states is much more irregular than the two-magnon example, and --- as we could convince ourselves in Section \ref{QFT} --- the degeneracy of the multi-trace operators amongst themselves is apparently not an obstacle.

\section{Hexagons and twist} \label{Hexagons}

In the spectrum problem, the full set of excitations is treated as follows:
\beq
\phi^{aa'} \rar \phi^a \bar \phi^{a'}, \quad \psi^{\alpha a'} \rar \psi^\alpha \bar \phi^{a'}, \quad \bar \psi^{a \dot \alpha} \rar \phi^a \bar \psi^{\dot \alpha}, \quad D^{\alpha \dot \alpha} \rar \psi^\alpha \bar \psi^{\dot \alpha}
\eeq
where $a \, = \, 1,2, \, a' \, = \, 3,4$ and $\alpha, \, \dot \alpha$ are two-component spinor indices. The \emph{left excitations} $\{\phi^a, \, \psi^\alpha\}$  and the \emph{right excitations} $\{\bar \phi^{a'}, \bar \psi^{\dot \alpha}\}$ transform under separate $PSU(2,2)_{L,R}$ groups that share a central extension. The left and the right excitation from any two-index scalar, fermion, or derivative put on the chain are given the same rapidity. The excitations now scatter separately on both chains with a 10 component $S$ matrix derived in \cite{psu22} times a phase factor \cite{bes}. Every such tensor product of $S$ matrices is divided by a single copy of \eqref{SU2mat}.

One recovers the simpler prescription we had given for the $SU(2)$ sector when only scalars of the same type are considered: on the left as well as on the right chain solely the component $A$ of the complete matrix \cite{psu22} is solicited, and this is exactly equal to \eqref{SU2mat} at tree level. Normalising by one more $A$ element we recover what we had before because the phase factor goes to unity at tree level.

The idea of the hexagon approach \cite{BKV} for three-point functions owes much to the earlier paper \cite{romuZoli}, in which a string theory three-point interaction was regarded as $\langle \mathrm{in} | \cO | \mathrm{out} \rangle$. The authors were trying to impose form factor axioms as is done in two-dimensional physics \cite{Karowski}. In \cite{BKV} the three-vertex is sliced through the middle resulting into two hexagonal patches. Three of these edges are string pieces, or in the dual field theory picture one half of a single-trace state (i.e. of a spin chain as we learned in the last section). These are called \emph{physical edges}. Between them lie \emph{virtual edges} that can in the field theory be understood as the bunches of configuration space propagators stretching between the three operators at tree level. In \cite{BKV}, axioms for these hexagonal patches were not only formulated but also solved.

\begin{center}
\includegraphics[height = 2.5 cm]{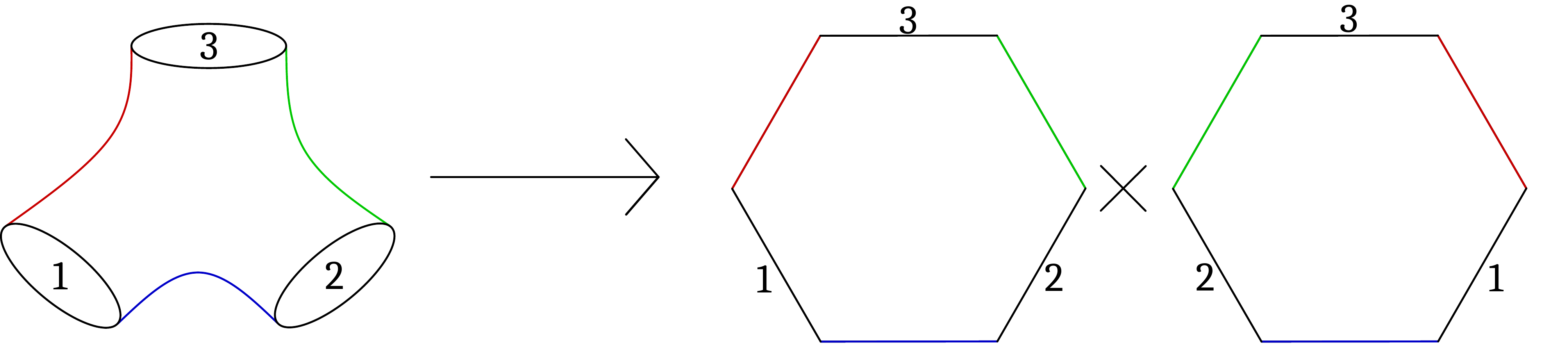}
\vskip 0.1 cm
\textbf{Figure 1}: splitting a three-point function into two hexagons
\end{center}
\vskip 0.2 cm
The resulting prescription for the computation of a structure constant is as follows:
\begin{itemize}
\item Find the Bethe solutions desribing the three selected primary operators.
\item Cut each Bethe state into two halves corresponding to its portions on the two hexagons.
\item Move all excitations to the same edge of the hexagon by \emph{crossing transformations}.
\item Split each two-index field or derivative from an operator into a left and a right magnon. 
\item Scatter only on the left or only on the right chain with the 10 component $PSU(2,2)$ invariant $S$ matrix \cite{psu22} supplemented by an adapted phase factor \cite{BKV}.
\item Use a certain contraction rule to put back together magnons with the same rapidity.
\end{itemize}
 The various hexagon scattering amplitudes needed for our purposes are appended in an ancillary file. We will not delve into their derivation; the techniques are well-explained in \cite{BKV}. Further comments relevant in the present context can also be found in \cite{cushions,colourDressed}. 

Nonetheless, we need to explain two features of the construction, the first of which is the \emph{twisted translation} of the vacuum $Z(0)$ to the rotated field
\beq
\hat Z(a) \, = \, Z + a \, (Y-\bar Y) + a^2 \bar Z \label{twistedT1}
\eeq
at some position $a$ along a real line. Likewise, for the \emph{longitudinal} magnons $Y, \, \bar Y$ one has
\beq
\hat Y \, = \, Y + a \, \bar Z \, , \qquad \hat {\bar Y} \, = \, \bar Y - a \, \bar Z \, . \label{twistedT2}
\eeq
In constructing structure constants from elements of the Bethe ansatz it is assumed that the three operators are located at $a \, = \, 0, \, 1, \, \infty$. 
To be able to compare we must project the propagators of free field theory according to \eqref{twistedT1} and \eqref{twistedT2}. For the \emph{transverse} $X$ nothing happens because it is not involved in the twisted translation:
\beq
\langle X(a_1) \bar X(a_2) \rangle \, = \, \frac{1}{(a_1-a_2)^2}
\eeq
Here we have omitted the colour indices and the normalisation factor $-1/(2 \, \pi)^2$ that decorates every massless configuration space propagator. There is no off-diagonal propagator tying $X$ to any of the other scalars so that $X$ amplitudes will only exist if there are equal numbers of $X$ and $\bar X$ fields. In fact, due to the contraction prescription and the crossing rules this is also valid on a single hexagon. Norm-like hexagon amplitudes involve an $SU(2)$ sector operator built from $X$'s at one point, the same operator at a second point though realised with $\bar X$ scalars, and a unit operator at the third end of the three-point function, cf. \cite{BKV}. 

On the hexagon the conjugate particle to $Y$ is $Y$ itself, again owing to the crossing rules in combination with the contraction prescription. Strangely enough, one can apparently insert the same operator at two ends of the three-point function and unity at the third to compute norms with the \emph{longitudinal} $Y$ magnons \cite{colourDressed}.

The \emph{effective scalar propagators} relevant to this exercise are (fields at $a_1,a_2$):
\beq
\langle Z Z \rangle \, = \, 1 \, , \qquad \langle Z Y \rangle \, = \, \frac{1}{a_2-a_1} \, , \qquad \langle Y Z \rangle \, = \, \frac{1}{a_1-a_2} \, , \qquad \langle Y Y \rangle \, = \, 0 \, . \label{yRules}
\eeq
In computing norms with these rules there are apparently no problems with primary single-trace states; the projected field theory results and the hexagon amplitudes agree. Yet, the last rule in \eqref{yRules} has an effect that will be problematic for our purposes:  we saw in Section \ref{QFT} that the mixing of the operator $(7,3)^+$ involves the descendent (5,3). Forming two-point functions using two identical operators realised from $\hat Z, \, \hat Y$ fields we calculate
\beq
\langle (5,2) (5,2) \rangle \, = \, N (N^2-1)(N^2-4) \, , \qquad \langle (5,3) (5,3) \rangle \, = \, 0 \, . \label{nor5253Y}
\eeq
The way the operators are defined in Section \ref{QFT} the two norms must be equal in field theory; a zero result would be non-unitary. In the hexagon picture, the twisted translation is the reason for this quirky accident because in the second case a $\langle Y Y \rangle$ contraction cannot be avoided. The issue will not only affect the (5,3) state but indeed any $SU(2)$ sector operator beyond \emph{half filling}. These are always descendents, but as the example shows they will occur in higher-trace admixtures to single-trace primaries. In this situation we should unfortunately expect a breakdown of the integrability picture we are trying to develop, or at the very least we have to accept that all $Y$ correlators cease to be norm-like despite of the interpretation of $Y$ as its own conjugate. The problem is a new type of \emph{finite size effect}. 

Second, the notion of an \emph{entangled state} allows us to split Bethe states \cite{tailoring,BKV}. This is best explained on the simplest non-trivial example: let us assume there are two magnons and bringing these from one hexagon to the other we would cross a bunch of $l$ propagators. The splitting is accomplished by:
\begin{eqnarray}
&& \psi(\{u_1,u_2\}) \rar \label{entangled} \\
&& \ \ \psi_1(\{u_1,u_2\}) \psi_2(\{\}) - e^{i \, p_2 \, l} \psi_1(\{u_1\}) \psi_2(\{u_2\}) - e^{i \, p_1 \, l} S_{12} \psi_1(\{u_2\}) \psi_2(\{u_1\}) + e^{i \, (p_1 + p_2)} \psi_1(\{\}) \psi_2(\{u_1,u_2\}) \nonumber
\end{eqnarray}
Higher cases work analogously. Like in the normalisation condition \eqref{normWave} it crucial to adhere to the ordering in which the magnons are originally put on the first few sites of the Bethe state we want to split. An $S$ matrix must be inserted for every magnon overtaking another one upon jumping to the second portion of the state. Last, the process may be iterated splitting the second portion again to arrive at a triple \emph{partition} etc. 

When calculating a structure constant with more than one renormalised operator the questions arises whether the result depends on the relative starting points of the partitions. For instance, for two non-trivial operators in a three-point function we could start both on the same hexagon or on different ones, also we can flip back and front of the figure. It is the Bethe equations and the momentum constraint that guarantee consistency: if we started the partitioning in \eqref{entangled} on the other hexagon, the magnons would transverse the other edge of width $L-l$. Keeping track of 
$\psi_1(\alpha) \, \psi_2(\bar \alpha)$ we find the conditions
\beq
e^{i \, p_1 \, l} \, S_{12} \, = \, e^{i \, p_2 \, (L-l)} \, , \qquad e^{i \, p_2 \, l} \, = \, e^{i \, p_1 \, (L-l)} \, S_{12}
\eeq
which reduce to the Bethe equation for the particle with $p_1$ upon putting $p_2 \, = \, - p_1$.

Seeking regular solutions of the Bethe equations at given $L, \, n$ one has as many equations as unknown rapidities. Due to the high order of the equations (written as polynomials) there will be several solutions, but the solution set is always discrete. The momentum constraint selects the cyclic cases, so those that are able to describe single-trace operators. In deriving the Bethe solutions for our $(2,1), \, (5,3), \, (6,3)^e$ solutions with twist we have used the momentum constraint in that vein: it is violated at $\ep \, \neq \, 0$ but its leading order singles out the desired solution. Indeed, multiplying all Bethe equations we find
\beq
e^{i \, n \, \ep} \, e^{i \, L \sum_{j=1}^n p_j} \, = \, 1 \qquad \Rightarrow \qquad e^{i \sum_{j=1}^n p_j} \, = \, e^{- i \, \frac{n}{L} \, \ep} \, .
\eeq
The hexagon itself is an off-shell object like the Bethe wave function prior to imposing cyclicity. On the other hand, we might guess that the twist of the Bethe equations should appear in building the entangled state as
\beq
e^{i \, p \, l} \rar e^{i \, n \, \ep} \, e^{i \, p \, l} \, .
\eeq
Let us modify the argument about the partition independence of the two-magnon entangled state. We define that any magnon crossing the edge of width $l$ picks up $n_A$ units of twists and any magnon crossing the edge of width $L-l$ should acquire $n_B$ units of twist. Equating the coefficients of $\psi_1(\{u_1,u_2\}) \psi_2(\{\})$ in the two situations we find can fix $n_A$, looking at $\psi_2(\{\}) \psi_1(\{u_1,u_2\})$ we determine $n_B$. The result is
\beq
n_A \, = \, \frac{l}{L} \, , \quad n_B \, = \, 1 - \frac{l}{L} \, .
\eeq
The coefficients of $\psi_1(\{u_1\}) \psi_2(\{u_2\}), \, \psi_1(\{u_2\}) \psi_2(\{u_1\})$ are then automatically partition independent. Also in a range of cases with more edges (so multiple partitions) and more magnons we could confirm that $n_i \, = \, l_i/L$ (where $\sum l_i \, = \, L$) assures partition independence. This solution realises $\sum n_i \, = \, 1$ as expected from the Bethe equations. On the other hand, $n_i$ is given different values at the opposite ends of an edge connecting operators of unequal length. Last, the outcome $n \, = \, l/L$ is perhaps not a surprise --- not catering e.g. for some jump in the twist at a certain point of the chain we have implicitly assumed an even distribution of the twist.

Unfortunately, this elegant manner of regularising the hexagon amplitudes does not reproduce the results of tree level field theory, cf. Section \ref{descCor} on the descendent correlators relevant to the $(7,3)^+$ mixing problem. In trying to design a viable scheme one can develop some fantasy: should the $n$'s around a chain really sum up to 1? Should $n$ depend on the edge width? On the same edge, do we have the same $n$ at either side? Should we put in a twist factor also for a magnon from an operator with a straight, so untwisted solution? Should the twist factors be more general functions?

For the time being we dropped the last idea because experiments with twist factors like $e^{i \, p \, \ep}$ do not seem to yield a suitable series expansion. For transverse magnons $X, \bar X$ the following principles proved helpful: 
\begin{itemize}
\item Whenever a magnon comes from a solution with twist $e^{i \ep}$ in the Bethe equation we associate extra factors $e^{i \, n_j \, \ep}$ to the edges crossed. So for untwisted Bethe solutions these extra factors are trivially 1. Below we will occasionally scale $\ep \rar c_0 \, \ep$ in both, the Bethe equations/solutions and the extra factors.
\item We will be pragmatic and put in such factors everywhere, not necessarily assuming that they are equal at both ends of the same edge. The condition that all $n$'s of any one operator must add up to the twist in its Bethe equation is then usually empty, because the last edge is not crossed in forming the partitions.
\item We impose consistency conditions: the independence of scaling the twist of any operator by $c_0$, and (where that becomes relevant) the existence of the two-point limit in calculating two-point functions of double-trace operators.
\item We allow $\{n_j\}$ to be specific to the partitioning chosen on a given diagram. For fixed partitioning we rather look for solutions independent of the edge width.
\end{itemize}
For the transverse magnons this program is quite successful: the $n$'s are usually 0 and sometimes 1; we emphasise that the non-trivial case does occasionally appear. The coefficients at the ends of the same edge should apparently be equal. Therefore, the condition $\sum n_i \, = \, 1$ around a chain does yield a helpful and consistent constraint from time to time. Unfortunately, as the aforementioned problems with the interpretation of all $Y$ correlators might have suggested, nothing works to plan for longitudinal magnons, cf. Appendix C.

Let us illustrate the procedure and its problems on the example of the leading $N$ norm of the $(5,3)$ operator: we use the original hexagon construction for a three-point functions on the sphere with two three-magnon operators and an identity. Let the non-trivial operators both be (5,3), once realised with $X$ magnons and once with $\bar X$ excitations. The fact that the two Bethe solutions are equal makes the computation run into the particle creation poles $1/(u-v)$ with $u$ a rapidity of the first operator and $v$ one of the second. These poles have to cancel between the various partitions. In practice, factoring them out of the complete amplitude is quite hard already for two sets of three off-shell rapidities. Introducing twist factors on the edges further complicates the issue. On the other hand, we can use the twist $\ep$ as a regulator if we rescale $\ep \rar c_0 \, \ep$ for the second operator. The particle creation poles will then appear as $1/(1-c_0)$. Organising the entangled states such that the magnons from both operators are shifted over the common width 5 edge we calculate the hexagon amplitude 
\beq
\cA \, = \, \frac{20 \, i \, c_0 \, \ep^2 (1 - c_0 - n_1 + c_0 \, n_2) (1 - c_0 + n_1 - c_0 \, n_2) (4 - 4 \, c_0 + 5 \, n_1 - 5 \, c_0 \, n_2)}{(1 - c_0)^3 (a_1 - a_2)^6} + O(\ep^3) \, . \label{norm53first}
\eeq
For every operator, this is to be divided by the root of the same phase factor and Gaudin determinant that we have encountered in the normalisation of the Bethe state given in \eqref{normWave}. The factor $\sqrt{\prod_i u_i^2 + 1/4}$ is not needed in the hexagon picture. To leading order in $\ep$ we compute $i$ for the phase and for the Gaudin norm $80 \, \ep^2$ or $80 \, c_0^2 \, \ep^2$ at the first and second point, respectively. We should thus normalise by $80 \, i \, c_0 \, \ep^2$.

Therefore, we demand
\beq
\frac{\partial \, \cA/c_0}{\partial \, c_0} \, = \, 0 \, . \label{cIndep}
\eeq
to guarantee a well-defined coincidence limit $c_0 \rar 1$. This imposes $n_1 \, = \, n_2$ for the coefficients at opposite ends of the edge. With that
\beq
\cA \, = \, \frac{20 \, i \, c_0 \, \ep^2 (1 - n_1)(1+n_1)(4 + 5 \, n_1)}{(a_1-a_2)^6} \, , \label{norm53}
\eeq
whereby the desired result is obtained putting $\na \, = \, 0$. This looks as if the regulator is not needed at all. It is instructive to repeat the exercise building the two entangled states starting on the same hexagon. To this end, let us move the magnons of the second operator over the corresponding width 0 edge. Now, condition \eqref{cIndep} reads $n_2 \, = \, 1 - n_1$. With that \eqref{norm53} is reproduced. Next, let us move the magnons of both states over the respective width zero edges. Instead of formula \eqref{norm53first} we find the same with $n_j \rar 1 - n_j$. The consistency reads again $n_1 \, = \, n_2$ and to reproduce the desired outcome we must set $n_1 \, = \, n_2 \, = 1$.

In conclusion, even if the twist is invisible organising the partitions as we first did, there is one unit of twist bringing each magnon once around the chain. This is lumped on the edges of width 0, so away from the edge connecting the two operators with equal rapidities.

Repeating the exercise for two equal operators with $Y$ magnons we obtain
\beq
\cA \, = \, \frac{20 \, i \, c_0 \, \ep^2 (1 - c_0 + n_1 - c_0 \, n_2) (4 - 4 \, c_0 + 5 \, n_1 - 5 \, c_0 \, n_2) (c_0 \, n_1 - n_2 + n_1 n_2 - c_0 \, n_1 n_2)}{(1 - c_0)^3 (a_1 - a_2)^6} \, .
\eeq
The independence of $c_0$ requires $n_1 \, = \, n_2$ as before, with which
\beq
\cA \, = \, \frac{20 \, i \, c_0 \, \ep^2 (1 - n_1)(1 + n_1)(4 + 5 \, n_1) \, n_1}{(a_1 - a_2)^6} \, .
\eeq
As for the transverse magnons, the other patterns of partitioning yield the same upon successively mapping $n_j \rar 1 - n_j$. Our observations about the location of the twist carry over.

Interestingly, the amplitude with longitudinal magnons is the one for transverse magnons times another factor $n_1$. Clearly, $n_1 \, = \, 0$ consistently reproduces the unwanted field theory result. Unfortunately, there is no real value for $n_1$ yielding $80 \, i \, c_0 \, \ep^2$. 

Finally, we recall that the hexagon amplitudes do not contain colour factors \cite{colourDressed}.

\section{Hexagon tessellations for higher-point functions} \label{tessel}

In \cite{cushions,shotaThiago1,colourDressed} it was argued that the hexagon operator can be used to compute higher-point functions of the $\cN \, = \, 4$ theory from integrability, too. For our purposes it will be good enough to stick to the restricted kinematics on the real line introduced above. In \cite{cushions, colourDressed} an orbital factor
\beq
v_{i;jk} \, = \, \frac{1}{a_i-a_j} - \frac{1}{a_i-a_k} \label{aVec}
\eeq
per scalar magnon on edge $i$ was included into every hexagon. These labels refer to the physical edges, and in the turning sense of the hexagon $j(k)$ are to the left(right) of $i$.  Since $v_{i;jk}$ is antisymmetric under $j \, \leftrightarrow \, k$ it flips sign switching from the first hexagon in Figure 1 to the second which has the opposite orientation. This explains the minus sign in \eqref{entangled} for transferring a magnon from the first portion of the entangled state to the second. Including the position vectors into the hexagons we have to write a plus sign instead. 

The claim is then that any triangulation of a punctured Riemann surface can be used to compute the contribution of a given tree level graph, if the latter can be drawn along the edges of the tiles.

\begin{center}
\includegraphics[height = 4 cm]{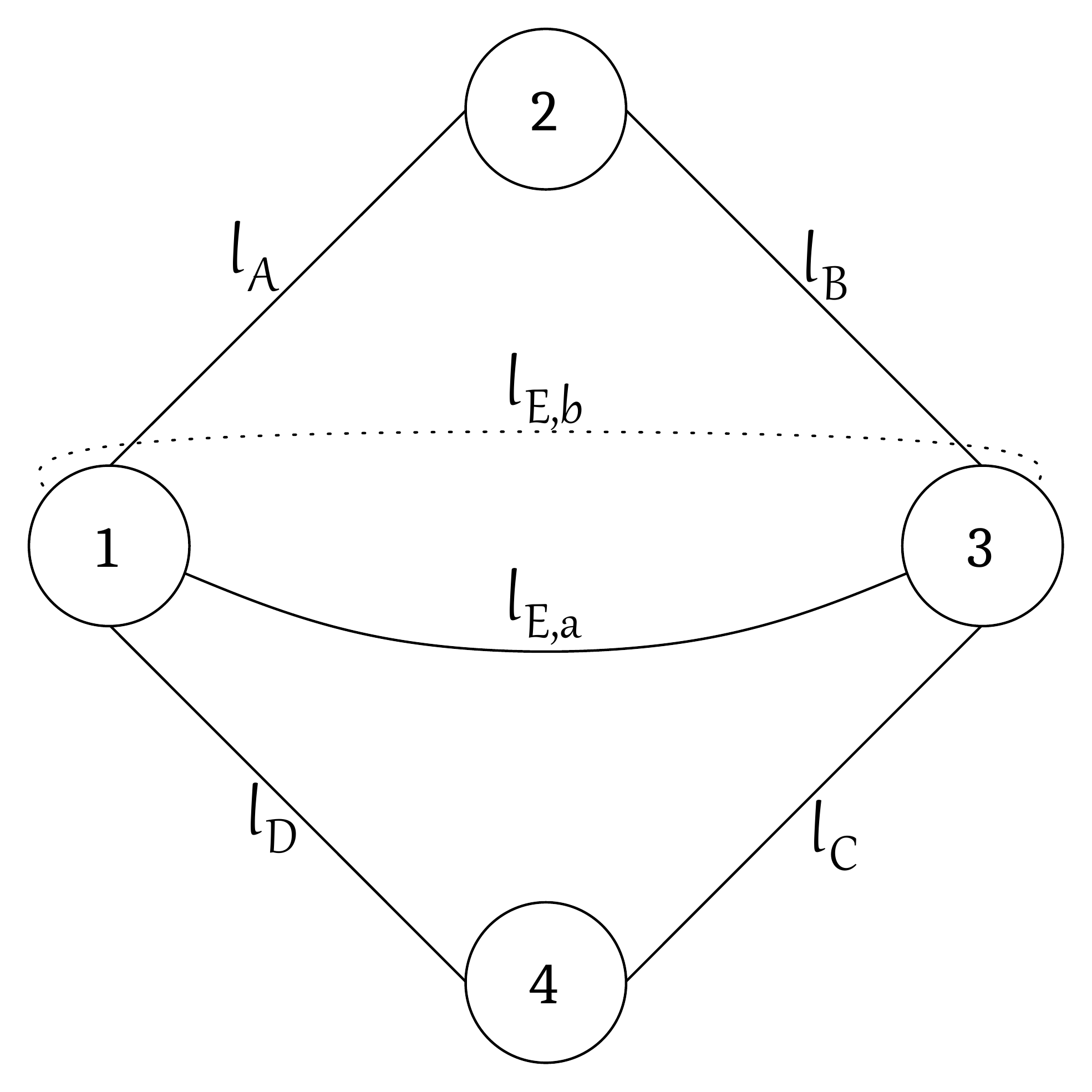}
\hskip 1 cm
\includegraphics[height = 4 cm]{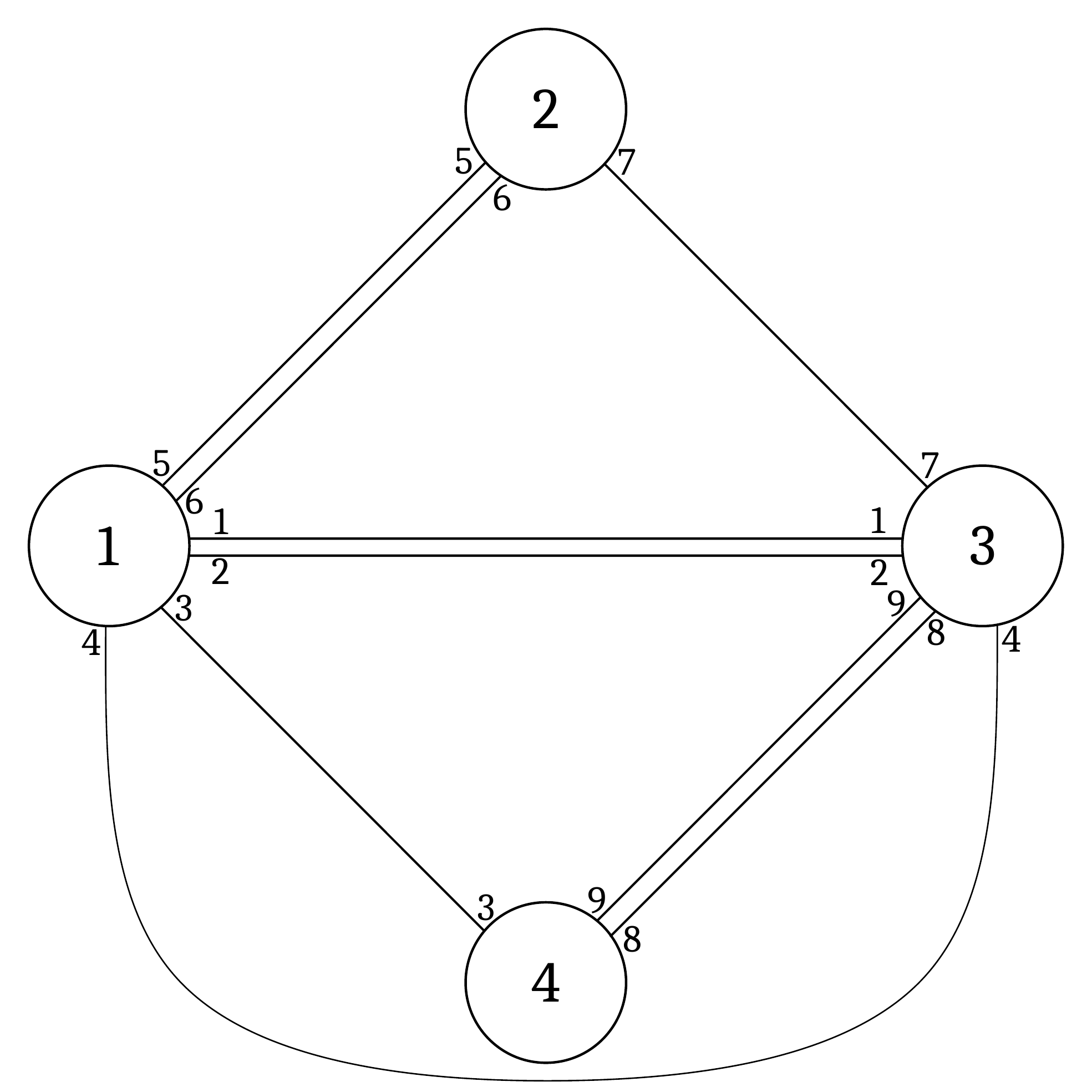}
\vskip 0.1 cm
\textbf{Figure 2}: a four-point diagram and the associated colour graph for edge widths $\{2,1,2,1,2,1\}$
\end{center}

Consider the four-point graph in the left panel of Figure 2 with two single-trace operators of length 6 at the punctures 1 and 3 and two operators of length 3 at the two other points 2,4. In the right panel we have opened up the graph on the plane to make it simpler to draw the seven propagators composing a planar graph on the sphere. We can translate the sample \emph{ribbon graph} on the right panel into the colour factor
\beq
\Tr(T^1T^2T^3T^4T^5T^6) \, \Tr(T^2T^1T^7T^4T^8T^9) \, \Tr(T^6T^5T^7) \, \Tr(T^3T^9T^8) \label{sampleRibbon}
\eeq
where $T^i$ is a short for $T^{a_i}$. Note that the indices from lines connecting two operators (above 1,2) occur anticyclically within the second trace when they are cyclically positioned on the first. We can stratify the contributions to free field theory correlators by such colour factors. Their coefficients in the tree QFT results with effective propagators  are faithfully reproduced by hexagon amplitudes with the right set of edge widths \cite{cushions, colourDressed}. The latter should neither depend on the tiling nor the way the entangled states are organised. 

We will try to reproduce double-trace/double-trace contributions from the diagram in Figure 2, Panel 1. As a first example, consider two $(63)^e(30)$ operators. As above, the single-trace parts of length 6 are placed at points 1,2 and the length 3 parts at points 3,4 with the intention to take a coincidence limit $a_3 \rar a_1, \, a_4 \rar a_2$. As usual, for normal-ordered operators there should be no self-contractions. Therefore we are looking for tree graphs of the form of an \emph{empty square} with $l_{E,b} \, = \, 0 \, = \, l_{E,f}$. There are the four possibilities $\{l_A, \, l_B, \, l_C, \, l_D\} \, = \, \{6,0,3,0\}, \, \{5,1,2,1\}, \, \{4,2,1,2\}, \, \{3,3,0,3\}$. The ribbon graph for the third of these is illustrated on the right panel of Figure 2.

To handle the particle creation poles we scale the twist of the second operator by $c_0$ as explained above. The $(6,3)^e$ operators behaves like any other primary state: potential twist factors in the entangled state do not even appear in the $O(\ep^0)$ term of the normalised hexagon amplitudes. For both, transverse and longitudinal magnons we obtain the results $1, \, 1/2, \, 1/6, \, 0$, respectively. For comparison, omitting the space-time factor, free field theory yields $N^9 - 10 \, N^7, \, 9 \, N^7, \, 3 \, N^7, \, 0$, respectively, at the relevant orders in $N$. These results are also valid for both, transverse and longitudinal magnons. As observed in \cite{colourDressed}, the connected part of the hexagon amplitudes needs to be scaled up by a factor $\sqrt{L_1 L_2 L_3 L_4} \, = \, 18$ to obtain a match. Note that the product of the sphere colour factors in the disconnected part is $N^9 - 9 \, N^7 + \ldots$ The discrepancy is accounted for by the torus part of the $(6,3)^e$ two-point functions times the sphere part of the $(3,0)$ two point-function and vice versa, cf. Appendix B.

Within the hexagon framework we can draw another class of planar tree graphs on the sphere: placing the two length 6 operators at points 1 and 3 of the diagram, and the two length 2 operators at points 2 and 4, we can put $l_A \, = \, 0 \, = \, l_C, \, l_B \, = \, 3 \, = l_D$. The coincidence limit is now $a_2 \rar a_1, \, a_4  \rar a_3$. Below, we will refer to this topology as \emph{belt around the belly}. Now, the three propagators connecting the length 6 operators can be distributed around the equator of the sphere as $\{l_{E,b}, l_{E,f}\} \, = \, \{3,0\}, \, \{2,1\}, \, \{1,2\}, \, \{0,3\}$. Here, the $\{3,0\}$ and $\{0,3\}$ cases should both be equal to the $\{3,3,0,3\}$ case in the last paragraph. We need to take into account only one of the three realisations of this graph. Yet, generically the $\{2,1\}$ and $\{1,2\}$ cases would both be needed, cf. \cite{cushions}. This is in fact a feature of the tree field theory computation, too. In the case at hand these graphs all vanish, though.\footnote{The twist regulator could bring them back by the $n_{E,b}, n_{E,f}$ coefficients. We learn by comparison with field theory and the empty square case that these $n$ coefficients must vanish.}

The complete two-point functions of the $(9,3)_i^- + \ldots$ operators contain a cross term
$\langle (6,3)^e (3,0) * (7,3)^- (2,0) \rangle$. We compute separately for the two $(7,3)^j$ Bethe states and then form a difference. In field theory as well as in the hexagon picture the amplitudes neither depend on the twist regulator nor on which type of excitation is chosen. The possible tree diagrams are $\{6,0,2,1\}, \, \{5,1,1,2\}, \{4,2,0,3\}$. In the first case, the colour factor identically vanishes. Field theory yields $-2 \sqrt{3}, \, 0$ for the first Bethe state in the two other cases, $+2 \sqrt{3}, \, 0$ for the second.

Upon multiplication by $\sqrt{2 3 6 7}$ the hexagon amplitudes yield the same, on the condition that we reverse the sign of the amplitude for the first (7,3) Bethe state. We have no explanation for this rule. The \emph{belt} diagrams non-trivially add up to zero in a surprisingly non-trivial manner:
\vskip 0.2 cm
\begin{center}
\begin{tabular}{l|c} 
$(l_{E,b},l_{E,f})$ & hexagon tiling \\
\hline
(4,0), (0,4) & $\frac{1}{2 \sqrt{21}}$ \\[1 mm]
(3,1) & $-\frac{1}{4\sqrt{21}} - i \frac{\sqrt{5}}{12 \sqrt{7}}$ \\[1mm]
(2,2) & 0 \\[1 mm]
(1,3) & $-\frac{1}{4\sqrt{21}} + i \frac{\sqrt{5}}{12 \sqrt{7}}$ \\[1 mm]
\end{tabular}
\end{center}
The four colour factors can be distinguished by the ribbon structure but are identically equal upon evaluation. Again, a sign needs to be amended on the integrability side for $\langle (6,3)^e (3,0) * (7,3)^+ (2,0) \rangle$ to be absent and $\langle (6,3)^e (3,0) * (7,3)^- (2,0) \rangle$ to exist.

Last, at the leading order in $N$, the two-point functions of $(7,3)^j$ in free field theory as well as integrability (times 14 in the connected part) for the $\{7,0,2,0\}, \, \{6,1,1,1\}, \, \{5,2,0,2\}$ tree graphs have the matrix form
\beq
N^9 \, \left(\begin{array}{cc} 1 & 0 \\ 0 & 1 \end{array}\right), \qquad \ 8 \, N^7 \, \left(\begin{array}{cc} 1 & 0 \\ 0 & 1 \end{array}\right), \qquad \frac{20}{3} N^7 \, \left(\begin{array}{cc} 1 & 1 \\ 1 & 1\end{array}\right) \, .  \label{mixing73}
\eeq
Again, in the integrability picture an extra minus sign on one of the states is required to match field theory. Considering the \emph{belt} cases we see a cancellation of imaginary parts within the pairs $\{4,1\}, \, \{1,4\}$ and $\{3,2\}, \, \{2,3\}$ and the sum $\{5,0\} + \{4,1\} + \{3,2\} + \{2,3\} + \{1,4\}$ is as stated. Since there is no twist regulator at our command to control the particle creation poles we have calculated numerically to high precision shifting the second set of rapidities by a small increment. The error comes at the expected order so that it is easy to truncate appropriately and match by the exact results stated above.

Our $(7,3)^+$ mixing example includes double-trace/double-trace two-point functions of the (5,3)(2,0) and (5,2)(2,1) operators involving descendents. The situation here is more complicated in two ways: with transverse magnons we occasionally need the twist regulator with non-trivial $n$'s as shown in Section \ref{descCor}. The regularised $Y$ correlators in the integrability picture could (almost) fit the $X$ combinatorics
but hardly the results for longitudinal magnons themselves, which deviate from the $X$ correlators. This is so even for $\langle (5,2)(2,1) * (5,2)(2,1) \rangle$, which is not beyond half filling in any obvious way. Tables are given in Appendix C.

\vskip 0.5 cm
\begin{center}
\includegraphics[height = 4 cm]{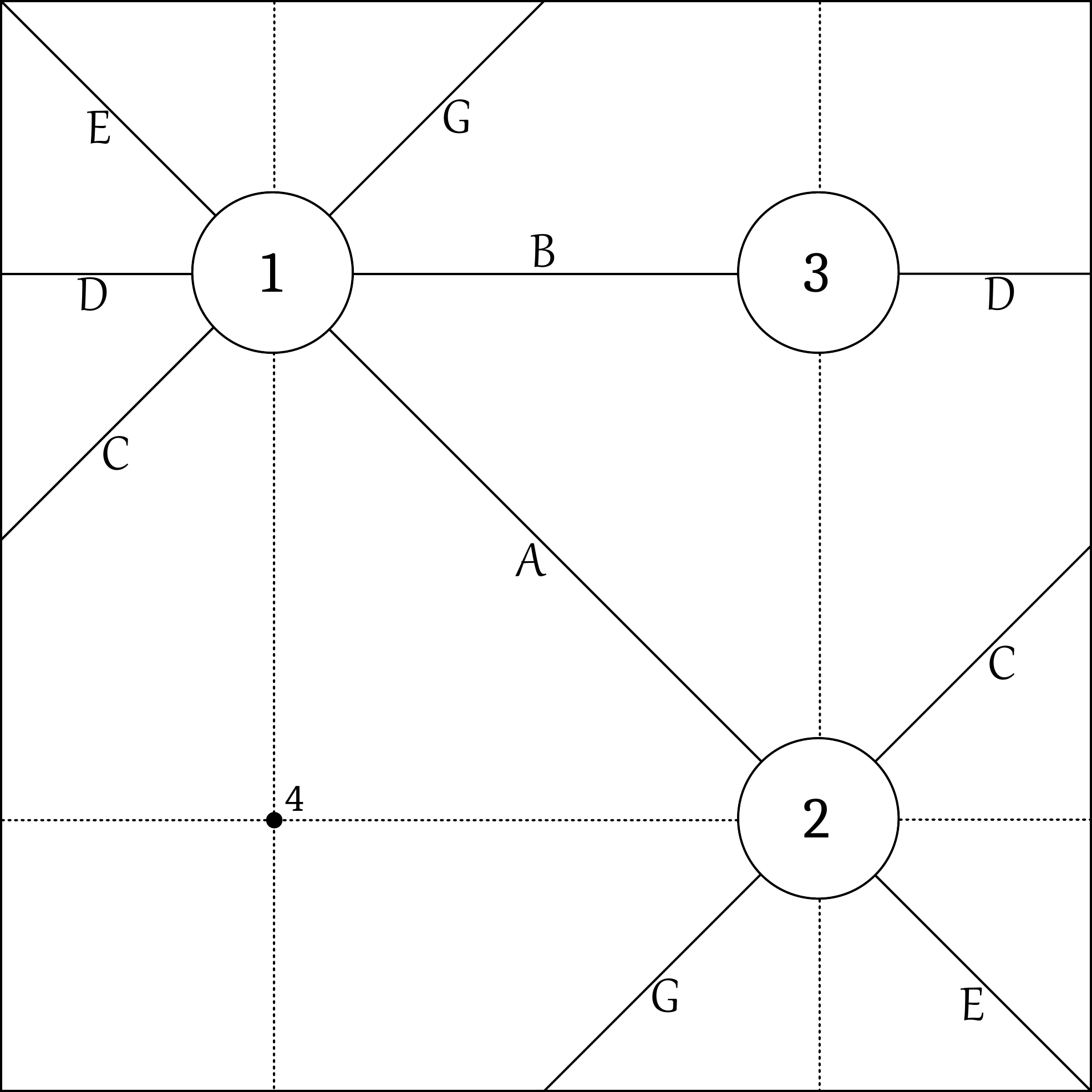}
\hskip 1.5 cm
\includegraphics[height = 4 cm]{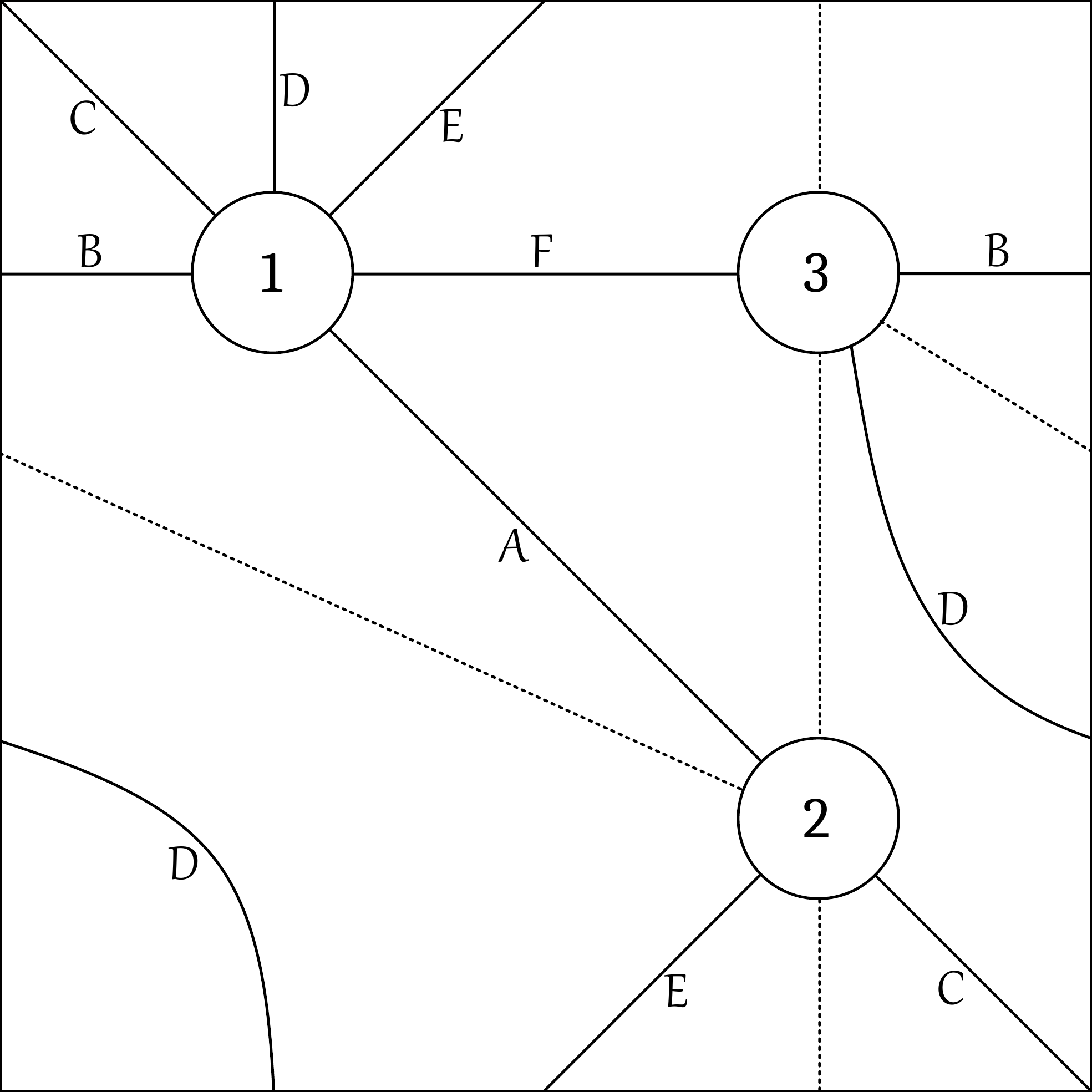}
\hskip 1.5 cm
\includegraphics[height = 4 cm]{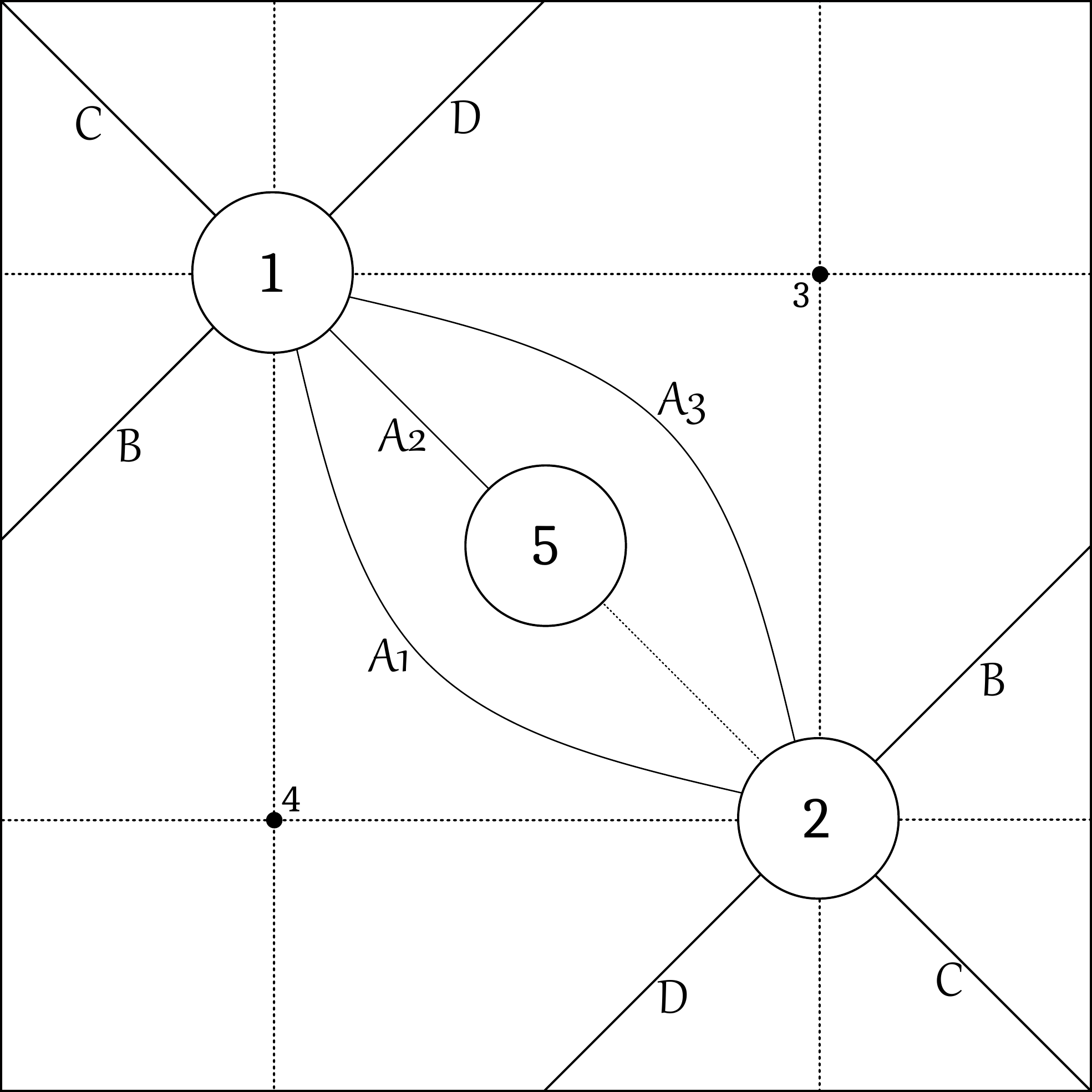}
\vskip 0.3 cm
\textbf{Figure 3}: diagrams for single-trace/double-trace correlators
\end{center}

\newpage

Next, we also wish to check the agreement between field theory and integrability for the single-trace/double-trace parts of correlators at leading order and on the torus. Here the first panel of Figure 3 may be employed, with a length 2 or 3 operator at point 3 and an identity insertion at point 4. For the primary operators we have prevalently calculated with longitudinal magnons; checks indicate that transverse magnons yield identical results, both in field theory and integrability. The space time dependence of the correlators is the expected factor $1/(a_1-a_2)^6$ whereby no problem with the coincidence limit can occur. We emphasise the following two points: first, when the two operators carrying excitations are linked by only one edge of the maximum width (or equivalents by cyclicity) the colour factor is of order $N^{L-1}$ with $L$ the greatest length. Yet, as these diagrams are connected (albeit \emph{extremal}) the same combinatorics applies as in the connected part of the double-trace/double-trace analysis above: the sphere part of the integrability computation needs to be scaled up by $\sqrt{L_1 L_2 L_3}$ like the true torus cases\footnote{The hexagon amplitude for the graph in Panel 2 of Figure 3 with edge widths $\{2,1,2,1,2,1\}$ overcounts the field theory result by a factor of 3 in this normalisation due to its threefold cyclic symmetry. The effect was first noticed in \cite{colourDressed}.}. Second, for the parity pairs we have to introduce an extra sign on $(9,3)^1, \, (7,3)^1$ again.

However, introducing the third non-trivial puncture into the game, two more types of colour factors can arise: first, the part of the double trace operator with the shorter length can lie inside one of the ribbons between the two longer operators, see Panel 2 of Figure 3. Second, in the $(9,3)^-$ problem we have mixing with $(6,3)^e (3,0)$. Obviously, we need to able to connect the length 9 and the length 3 operator with three single lines, which cannot be realised on the first two tilings. Here we have used the diagram in Panel 3.

The parity even $\langle (7,3)^+ * \cO_\perp \rangle$ correlator also requires the sign flip on the $(7,3)^1 \ldots$ hexagon amplitudes. Then everything works out with transverse as well as longitudinal magnons without any twist-like modification of the entangled states. We obtain
\beq
\braket{ (7,3)^+ * (5,3) (2,0) } \, = \, 2 \, \frac{\sqrt{10}}{\sqrt{3}} \, (N^6 - 5 \, N^4 + \ldots) \, = \, - \sqrt{2} \, \braket{ (7,3)^+ * (5,2) (2,1) } \, . \label{op73perp}
\eeq

Finally, we need single-trace/single-trace two-point functions on the sphere, the torus, and the double torus. In principle, tessellating the latter works in the obvious way and could be handled as described above. Yet, the workload quickly grows out of hand because there are relatively many distinct sets of edge widths. Consequently, we used a {\tt Mathematica} script to generate all Wick contractions, to classify these by the colour factors as in the right panel of Figure 2 or equation \eqref{sampleRibbon}, and to automatically embed the corresponding graph on the surface of the smallest possible genus. We have chiefly studied amplitudes for longitudinal magnons.

Where twist is not around to help with the particle creation poles we have regularised by a small shift of one set of rapidities as was already done to obtain \eqref{mixing73}. For the double torus computations we have used minimally 100 digits of precision and a shift of order $10^{-25}$ to arrive at 20 digits of precision in the output; we kept the precision fairly low to save on computer time.
\vskip 0.1 cm
\begin{center}
\includegraphics[height = 6 cm]{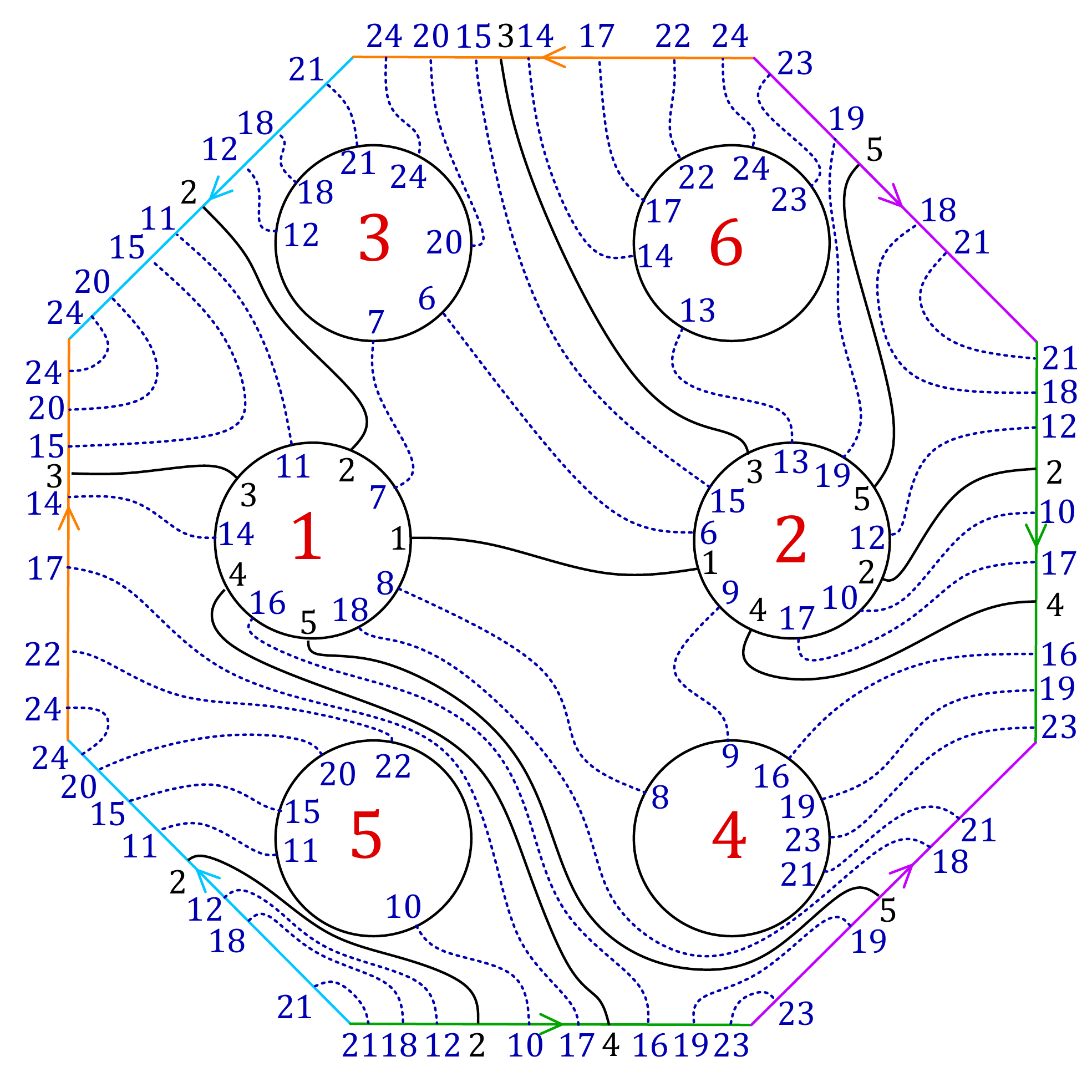}
\end{center}
\textbf{Figure 4}: a tessellation of the double torus with a colour ribbon graph for the $\braket{ (5,2) * (5,2) }$ correlator marked on it. The length five operators are at points 1 and 2. Four identity operators were added to achieve a triangulation. Actual propagators are depicted as solid lines, auxiliary edges of width zero as dotted lines. \\

The simplest example in which a genus 2 diagram contributes is the norm of the (5,2) operator. Due to its low length the operator does not receive admixtures so that all diagrams can be embedded on the double torus. Integrability supplemented by colour factors exactly reproduces the norm stated in \eqref{nor5253Y}.

\section{Twisting $\langle XXX \, \bX\bX\bX \rangle$} \label{descCor}

\subsection{$\mathbf{\langle(5,3)(2,0)*(5,3)(2,0)\rangle}$}

Like in the $(6,3)^e(3,0)$ two-point function we put the single-trace operators carrying excitations at points 1 and 2 and their complements (2,0) in the double trace products at points 2 and 4. The coincidence limit $3 \rar 1, \, 4 \rar 2$ is even trivial for the \emph{empty square} diagram with $\lh \, = \, 0 \, = \, \lv$ in Figure 2, Panel 1. To regularise the particle creation poles we rescale once again the regulator at the second point, i.e. we employ $c_0 \, \ep$ there. In the $\ep \rar 0$ limit there are infinite rapidities so that the situation resembles the norm example at the end of Section \ref{Hexagons}. The $c_0$ independence condition \eqref{cIndep} forces
\beq
\na \, = \, \naa \label{beforeTheTable}
\eeq
which is common to all four cases in the table below, even prior to the two-point limit. Omitting the obvious space time factor $1/(a_1-a_2)^6$:
\vskip 0.2 cm
\begin{center}
\begin{tabular}{l|c|l} 
$\la\lb\lc\ld$ & QFT & hexagon tiling \\
\hline
5020 & 1 & $\phantom{10 *} \ \ \frac{1}{4} \, (1 - \na)(1 + \na)(4 + 5 \, \na)$ \\[1mm]
4111 & 4 & $10 * \frac{1}{20} (1 + \na) (2 - 5 \, \na) (4 + 5 \, \na)$ \\[1mm]
3202 & 1 & $10 * \frac{1}{20} (2 + 5 \, \na)(1 - 8 \, \na - 5 \, \na^2)$ \\
\end{tabular}
\end{center}
In the field theory results we have indicated the leading $N$ coefficient only. Clearly, for a match we must put $\na \, = \, 0$. For a test we have calculated the $\langle (5,3)(2,0) * (5,3)(2,0) \rangle$ hexagon amplitudes also sending the magnons of the operator at point 2 over the edge $B$. Then $c_0$ independence implies $\na \, = \, 1 - n_{B,2}$ as in the norm example in Section \ref{Hexagons}. Turning the entangled state at point 1 nothing changes as long as the edge $A$ is crossed. On the other hand, if the magnons go over the edges $E_f,\, D, \, E_B$ we find the same with $\na \rar 1 - \nv - \nd - \nh$ and to match QFT one has to assign $\nv + \nd + \nh \, = \, 1$.  Hence we cannot resolve exactly where the twist has to go.

For the \emph{belt around the belly} version we put the two non-BPS operators at points 1 and 3 and the vacua at 2 and 4. On the equator lines of the diagram we have in principle four different indices --- one $\nh$ and one $\nv$ at each point. For brevity we have put all of these equal. The $c_0$ independence constraint is then empty. We organise the partitions such that the magnons are clockwise brought over the backwards equator edge first. The excitations thus never cross the two edges $B, \, D$ of non-vanishing length, so that their $n$ coefficients cannot appear. We actually find results without the $n_A, \, n_C$ either:
\vskip 0.2 cm
\begin{center}
\begin{tabular}{l|c|l} 
$\{\lh,\lv\}\la\lb\lc\ld$ & QFT & hexagon tiling \\
\hline
\{3,0\}0202 & 1 & $10 * \frac{1}{10} (1 + 9 \, \nh + 10 \, \nh^2)$ \\[1mm]
\{2,1\}0202 & 2 & $10 * \frac{1}{10} (1 - 3 \, \nh + 10 \, \nh^2)$ \\
\end{tabular}
\end{center}
Once again, we must drop the regulator. Note that the second line in the table shows the necessity of doubling the $\{2,1\}$ graph. Upon tracking the colour factors the field theory result actually contains both, a $\{2,1\}$ and a $\{1,2\}$ part while $\{3,0\}$ can only be realised in one way.

In the light of our previous experience an interesting experiment to make is to move the magnons of the left operator over the edges $D, \, E_b, \, A$ and those of the right operator over $C, \, E_f, \, B$. From \eqref{cIndep} we find the two constraints
\beq
\na + \nd \, = \, 1 - \nh - \nv \, , \qquad \nb + n_{C,3} \, = \, 1 - \nh - \nv \, .
\eeq
If we now assume $\nv \, = \, \nh$ we fall upon the results in the table whereby $\nh \, = \, 0$. Nonetheless, we have learned $\na + \nd \, = \, 1$ and $\nb + n_{C,3} \, = \, 1$ and so the twist exists, obeys the familiar sum rule from the Bethe equations. Once again, we cannot precisely localise it. Here it is distributed around the square frame, i.e. not on the lines connecting operators with identical rapidities. 

\subsection{$\mathbf{\langle(5,3)(2,0)*(5,2)(2,1)\rangle}$}

We will now put the (5,3) operator at point 1, its partner (2,0) at the opposite point 3, and the (5,2) primary at point 2 its partner, the vacuum descendent (2,1) at point 4 of the \emph{empty square} in Figure 2, panel 1. We will associate $n$ coefficients to the states at points 1 and 4; the (5,2) operator at point 2 is undeformed, so it seems logical to not associate it with $\ep$ at all.\footnote{In a more general ansatz one might artificially introduce twist into the (5,2) Bethe equations and drag such coefficients along.} At point 1 we organise the partition as before, while at point 4 we move the magnons clockwise over the edge $C$. Coefficients to be expected in the result are now thus $\na, \, \nd$ and $\nv$ at point 1 as well as $\nc$.

Again, we will distinguish the twist at point 4 from that at point 1 by a scale $c_0$ as one must be able to move both parameters independently. The amplitudes at $O(\ep^2)$ are polynomials of $c_0$ and the $n's$ over a denominator $2-c_0$. Clearly we should impose 
\eqref{cIndep} which yields
\beq
\nd \, = \, 2 \, (1 - n_{C,4}) \label{indep52522021}
\eeq
as a simple solution common to all three cases in the table below. Remarkably, had we assumed equal coefficients at both ends of the $D$ egde, so $\nd \, = \, n_{D,4}$, the twisted Bethe equation \eqref{twistedBethe} would have made us expect $\nd \, = \, 1 - n_{C,4}$. Reconciling this with the $c_0$ independence of the formula singles out $\nc \, = \, 1$, which will indeed be our result, see below. 

Employing \eqref{indep52522021} we obtain
\vskip 0.2 cm
\begin{center}
\begin{tabular}{l|c|l} 
$\la\lb\lc\ld$ & QFT & hexagon tiling \\
\hline
5020 & 0 & $\phantom{10 *} \ \, \frac{\sqrt{2}}{4} (1 + \na) (4 + 5 \, \na) (-1 + \nc)$ \\[1mm]
4111 & $ \sqrt{2}$ & $10 * \frac{\sqrt{2}}{40} (4 + 5 \, \na) (-5 - 5 \, \na + 6 \, \nc + 10 \, \na \nc)$ \\[1mm]
3202 & $ \sqrt{2}$ & $10 * \frac{\sqrt{2}}{20} (2 + 5 \, \na) (-2 + 3 \, \nc + 5 \, \na \nc)$ \\
\end{tabular}
\end{center}
for the leading $N$ parts. Demanding equality of field theory and normalised hexagon amplitudes simultaneously in all three cases is possible and uniquely determines
\beq
\na \, = \, 0 \, , \qquad \nc \, = \, 1 \, .
\eeq
It is also possible to keep the rapidities of the (5,2) operator undetermined during the $\ep$ expansion. The particle creation pole factors out upon which we can substitute the Bethe solution $u_2 \, = \frac{1}{2} \, = \, - u_1$. The $\ep$ expansion does not bring out $\na$ in this case whereas $c_0$ independence still creates the same constraint. We directly land on the three results in the table evaluated at $\na \, = \, 0$. With hindsight, this and the Bethe argument uniquely pin down the three amplitudes. In conclusion, this example requires introducing a single factor $e^{i \, c_0 \, \ep}$, namely when the magnon of (2,1) transverses the edge $C$ {\it independently of the edge width}. We could carry away a second message: 
there should be no unit of twist on lines connecting deformed operators.

Reorganising the partitions we can generate the transformations $\nd \rar 1 - \nh - \na - \nv$ and/or $\nc \rar 1 - n_{D,4}$ on the set of parameters. As expected by now, the content of the equations does not change. In particular, the second move allows us to drop all $n$ coefficients although the twist is inherently present. The example illustrates very well how the descendent amplitudes may collapse or come out wrong for some choice of partitions when no $n$'s are used.

In the \emph{belt} embedding we put the (5,3) operator at point 1, (2,0) at point 2, (5,2) at point 3 and finally (2,1) at point 4. The magnons at point 1 go over $\lh, \, \la, \, \lv$ as before and we also move that of (2,1) over edge $C$. Again, for simplicity we assume $\nh \, = \, \nv$. The $c_0$ independence constraint valid in both cases is
\beq
\nh \, = \, -\frac{1}{2} - \frac{\na}{2} + \nc \, ,
\eeq
with which:
\vskip 0.2 cm
\begin{center}
\begin{tabular}{l|c|l} 
$\{\lh,\lv\}\la\lb\lc\ld$ & QFT & hexagon tiling \\
\hline
\{3,0\}0202 & $\sqrt{2}$ & $10 * \frac{\sqrt{2}}{20} (1 + 5 \, \na +\nc + 5 \, \na \nc - 10 \, \nc^2)$ \\[1mm]
\{2,1\}0202 & $-2 \sqrt{2}$ & $10 * \frac{\sqrt{2}}{20} (-5 - 5 \, \na + 13 \, \nc + 5 \, \na \nc - 10 \, \nc^2)$ \\
\end{tabular}
\end{center}
Taking into account the doubling of the second diagram the two equations have the simultaneous solutions
\beq
\na \, = \, 1, \ \nc \, = \, 1 \qquad \lor \qquad \na \, = \, \frac{2}{5}, \ \nc \, = \, \frac{1}{2} \, .
\eeq
and so $\nh \, = \, 0$ or $\nh \, = \, -\frac{1}{5}$. In the second solution, the value of $\na$ suggests to try the homogeneous distribution of twist $n \, = \, l/L$ initially discussed. However,  once again that fails to yield the right results.

Aesthetically, the first solution seems more appealing --- also, it is more in line with what we have seen so far: the two sum rules we expect are $1 - \nc - n_{D,4} \, = \, 0, \, = \, 1 - \na - \nv - \nd - \nh$. Assuming $\nh \, = \, 0 \, = \nv$ (equal rapidities at $\ep \rar 0$) and $n_{D,j} \, = \, 0$ (the edge connects two operators carrying twist) we could have conjectured $\na \, = \, 1 \, = \, \nc$. We can hide the twist choosing the partitions such that the edges $A, \, C$ are avoided and putting all $n$'s to 0.

\subsection{$\mathbf{\langle(5,2)(2,1)*(5,2)(2,1)\rangle}$}

For the \emph{empty square} we put the two (5,2) operators at points 1,2 and the two vacuum descendents at 3,4. To the latter we assign twist $\ep, \, c_0 \, \ep$ to regularise their particle creation pole in the amplitude. We move both magnons over the $C$ edge. The (5,2) Bethe solution is undeformed so that we cannot rescale any epsilon here; we will substitute $u \, = \, \mp \frac{1}{2}$ for the first operator, expand in $\ep$, factor the amplitude, and finally put the rapidities of the second (5,2) operator on shell.  As expected, the amplitudes have a pole $1/(1-c_0)$ and regulator independence implies $n_{C,3} \, = \, \nc$. We find the table
\vskip 0.2 cm
\begin{center}
\begin{tabular}{l|c|l} 
$\la\lb\lc\ld$ & QFT & hexagon tiling \\
\hline
5020 & 1 & $\phantom{10 * \frac{1}{10}} (1 - \nc)$ \\[1mm]
4111 & 5 & $10 * \frac{1}{10} (5 - 6 \, \nc) $ \\[1mm]
3202 & 2 & $10 * \frac{1}{10} (2 - 3 \, \nc) $ \\
\end{tabular}
\end{center}
For a leading $N$ match it follows $\nc \, = \, 0$, so the twist is hidden on the edges $B, \, D$.

The \emph{belt around the belly} version has the two (5,2) operators at points 1 and 3 and the vacuum descendents at 2 and 4. The results are $n$ independent:
\vskip 0.2 cm
\begin{center}
\begin{tabular}{l|c|l} 
$\{\lh,\lv\}\la\lb\lc\ld$ & QFT & hexagon tiling \\
\hline
\{3,0\}0202 & 2 & $10 * \frac{1}{5}$ \\[1mm]
\{2,1\}0202 & 0 & 0 \\
\end{tabular}
\end{center}
Fortunately, this comes out right!

\section{Conclusions}

In this article we have studied the operator mixing in the so-called $SU(2)$ sector in $\cN \, = \, 4$ super Yang-Mills theory. Specifically, we have focused on multi-trace admixtures to single-trace operators in a number of examples with the aim of constructing a large $N$ expansion of the form $\cO^s + 1/N \, \cO^d + 1/N^2 \, \cO^t + \ldots$, where the subscripts refer to single-, double-, and triple-trace operators. 

As is well-known \cite{headrickPlefka,moreDegenerate,stauPlefChristi}, the degeneracy of the spectrum poses a problem: if a single-trace and a double-trace leading $N$ eigenstate have the same anomalous dimension, we naively find the sum and difference of the two operators as eigenstates, without an offset of $1/N$ in the relative coefficients. Hence the large $N$ expansion is not of the desired form. Degenerate perturbation theory can improve on this. It has also been suggested to look at the $\beta$ deformed theory \cite{beta} instead in which the degeneracy is lifted. The idea would be to define the states of the undeformed theory by the limit $\beta \rar 0$ \cite{whoBeta}. It would be interesting to address the operator mixing also in such cases; clearly hexagon tilings will still correctly reproduce the various overlaps.

Degeneracies are particularly frequent in the two-excitation spectrum. For operators with more excitations the eigenvalue problems have characteristic polynomials of higher degree whose roots rarely coincide with with those of lower cases. This is so in particular for so-called primary states, i.e. those that cannot be derived from others by $SU(2)$ raising. Adding in higher-trace admixtures we found that their leading $N$ anomalous dimensions should also differ from that of the single-trace state in question. Importantly, degeneracies within the relevant set of multi-trace operators are apparently irrelevant. Scanning the space of single-trace operators from length 2 to 10 with up to four excitations we found that the systematic $N$ expansion can be constructed in most cases, at least to the order indicated above.

An interesting feature of the higher-excitation spectrum is the existence of parity pairs, i.e. pairs of single-trace operators with degenerate leading $N$ anomalous dimension. In the exact planar limit we are free to choose any basis for such a $2 \, \times \, 2$ cell. One can always assume this basis to consist of the odd and even part under parity, here defined as the reversal of all colour traces. Parity is in fact strictly respected in the operator mixing, and the even and odd operators acquire distinct $1/N^2$ corrections to the common leading $N$ anomalous dimensions. 

To spare some work we studied three-excitation cases up to length 9 in which no triple-trace operators occur, and in which the mixing does not involve too many distinct operators. In order to compute two-point functions of the $N$ completed single-trace operators we obviously need to evaluate those of all their constituents, which we were able to reproduce  by hexagon tilings of the sphere, the torus, and the double-torus. In a coincidence limit, three-point functions on the sphere and the torus can be used to compute single-trace/double-trace mixing to the required order, and four-point functions on the sphere should reproduce the double-trace/double-trace contributions.

In a similar vein, the single-trace/double-trace (and vice versa) transitions caused by the one-loop dilatation operator can be captured by an overlap formula \cite{whoBeta} in the spin chain picture. For simplicity, we have rather imported the one-loop mixing matrices from field theory and then derived their eigenvectors by traditional means. A third ingredient is a similarity transformation by the (root of) the matrix of tree level two-point functions. This is  given by the very pieces that we did reproduce from integrability supplemented by colour factors \cite{colourDressed}. It would be worthwhile looking for a more systematic approach to the entire diagonalisation process.

Admittedly, the current technology constitutes at most a proof of principle because the integrability calculations are more cumbersome than tree level field theory.  Nonetheless, for primary operators --- including an exceptional case, which the Bethe ansatz misses unless twist is introduced --- the tessellation method works perfectly. Interestingly, we need to introduce an extra sign on one of the Bethe states describing each degenerate pair.

In our picture, multi-trace operators are products of single-trace eigenstates. As our $(7,3)^+$ example shows, higher-trace admixtures to primary states can have multi-trace admixtures in which one or more of the factors are descendents. The Bethe ansatz projects out descendent states, but like the aforementioned exceptional operator they may be brought back introducing twist into the Bethe equations.

The original hexagon construction \cite{BKV} as well as its application to higher-point functions \cite{cushions,shotaThiago1} cuts up the Riemann surface on which a graph can be drawn. In this process also the single trace operator viz Bethe state at a punture is cut into various pieces. As a consequence, we have to address the question as to how this can be made compatible with twist. For the single-trace/single trace and single-trace/double-trace two-point function we studied there is no need to alter the cutting rules to reproduce field theory from integrability. 

The situation changes w.r.t. the double-trace/double-trace two-point functions in our $(7,3)^+$ example: for transverse magnons in the terminology of \cite{BKV} an integrability/field theory match can be achieved using a number of empirical rules:
\begin{itemize}
\item We introduce a factor $e^{i \, \ep}$ into each Bethe equation for a descendent state. Bringing a magnon once around the entangled state at the corresponding puncture we should include a factor $e^{i \, n_j \, \ep}$ when the edge $j$ is crossed.
\item Homogeneously distributing the twist putting $n_j \, = \, l_j / L$ (with $l_j$ the width of the edge and $L$ the length of the operator) generically fails. Instead, the twist is inhomogeneously distributed depending on the tessellation and the position of the operators on it. Note that we did not have to include twist into our map from Bethe states to field theory operators which implies that the twist is lumped at one site of the chain.
\item In the examples studied, we cannot uniquely assign $n_j$  coefficients. Yet, around an entangled state they should add up to 1 as expected from the Bethe equations. If an edge connects two operators with twist, or two operators with Bethe rapidities degenerate at zero twist, the corresponding $n$ coefficient ought to vanish. 
\item Partition invariance --- the independence of the outcome of the way the Bethe states are cut --- is guaranteed in that different choices only lead to reparametrisations but do not alter the results.  
\end{itemize}
Imposing these rules of thumb the method does have predictive power. Yet, a more comprehensive study is clearly needed: for once we have not considered double or even higher descendents. Second, for higher-point functions with descendents at many punctures it is not clear that the rules above can be consistently imposed.

Furthermore, realising the $SU(2)$ sector with longitudinal magnons \cite{BKV} we have not found a way to reconcile integrability with field theory results for our double-trace/double-trace two-point functions involving descendents at both ends. The tree field theory results deviate form those for transverse magnons in these cases, which is an effect of the twisted translation built into the hexagon approach. Curiously, the hexagon amplitudes listed in Appendix C fall upon the expressions for transverse magnons. If not resolved, the issue causes a new type of finite size problem for integrability: a systematic large $N$ expansion cannot be dealt with when single-trace operators mix with descendents of shorter operators as factors of some multi-trace admixture.

The symmetry underlying the twisted translation is at odds with the existence of twist in this sector. It is therefore of vital interest to study for what other excitations of the complete $\cN \, = \, 4$ super spin chain similar difficulties arise. We might hope that a nested Bethe ansatz for the Dynkin diagram employed in \cite{beiStau,psu22} opens a way of removing the difficulties. In a putative hexagon approach to the $\beta$ deformed theory this point will be of central importance.

\section*{Acknowledgements}

B.~Eden is supported by Heisenberg funding of the Deutsche Forschungsgemeinschaft, grant Ed 78/7-1 or 441791296.  We are grateful to D.~le~Plat and T.~Mc~Loughlin for discussions about related material, and to C.~Kristjansen, A.~Spiering and T.~Mc~Loughlin for comments on the manuscript.

\section*{Appendix A: two-point functions up to $1/N^4$}

\subsubsection*{Exceptional operator at length 6}

The $(6,3)^e$ operator of equations \eqref{op63e}, \eqref{op63eInt} has the two-point function
\beq
\braket{(6,3)^e * (6,3)^e} \, = \, N^6 \left( 1-5\, N^{-2}+4 \, N^{-4}\right).
\eeq

\subsubsection*{Degenerate pair at (7,3)}

The Bethe states $(7,3)^j$ in \eqref{pair7} have overlaps
\begin{align}
\label{L7Norm1}
\braket{(7,3)^1 * (7,3)^1} \, = \, \braket{(7,3)^2 * (7,3)^2} &=\, N^7 \left(1-8\,N^{-2}+19\,N^{-4}+\cO(N^{-6})\right), \\
\label{L7Cross}
\braket{(7,3)^1 * (7,3)^2}&=\, N^7 \left(0+6\, N^{-2}-30\, N^{-4}+\cO(N^{-6})\right) \, .
\end{align}
From here it follows
\begin{align}
\label{L7Cross}
\braket{(7,3)^- * (7,3)^-} &=\, N^7 \left(1-14\, N^{-2}+49\, N^{-4}+\cO(N^{-6})\right) \, , \\
\label{L7Norm1}
\braket{(7,3)^+ * (7,3)^+} &=\, N^7 \left(1-\phantom{1}2\,N^{-2}-11\,N^{-4}+\cO(N^{-6})\right)\, . \nonumber
\end{align}
While we are done with the parity odd two-point function, we still need to discuss the effect of the admixtures to the parity even one. From \eqref{op73perp} we find
\beq
\braket{ (7,3)^+ * \cO_\perp } \, = \, - 2 \, \sqrt{5} \, (N^6 - 5 \, N^4 + \ldots)
\eeq
and adding the various results for transverse magnons (we drop the longitudinal case for now due to the difficulties displayed in Appendix C):
\beq
\braket{ \cO_\perp * \cO_\perp} \, = \, N^7 + N^5 - 26 \, N^3 +  \ldots
\eeq
From here we could straightforwardly compute the two-point functions of the two eigenstates \eqref{eigen73} in Section \ref{QFT}. However, the mostly single-trace and the mostly double-trace state are not orthogonal. As has been pointed out in the literature \cite{treeBusiness}, this can be mended by a similarity transform
\beq
M \rar S^{-\frac{1}{2}} \, M \, S^{\frac{1}{2}} \, , \qquad S_{ij} \, = \, \braket{ \cO_i \, \cO_j }_\mathrm{tree} \, .
\eeq
The root of the tree matrix $S$ has fairly complicated $N$ dependence. We resort to expanding up to NNLO in $N^{-2}$:
\beq
\sqrt{\frac{S}{N^7}} \, = \, \left( \begin{array}{ll} 1 - \frac{7}{2 \, N^2} - \frac{3}{8 \, N^4} &
  - \frac{\sqrt{5}}{N} + \frac{9 \, \sqrt{5}}{4 \, N^3} \\[1 mm] - \frac{\sqrt{5}}{N} + \frac{9 \, \sqrt{5}}{4 \, N^3} & 1 - \frac{2}{N^2} - \frac{15}{4 \, N^2} \end{array} \right) + \ldots \label{upRoot}
\eeq
The similarity transform maps $M$ as stated in \eqref{nonHerm} to
\beq
M \rar \left( \begin{array}{ll} 5 - \frac{5}{N^2} - \frac{5}{2 \, N^2} & -\frac{3 \, \sqrt{5}}{N} + \frac{3 \, \sqrt{5}}{4 \, N^3} \\[1 mm] -\frac{3 \, \sqrt{5}}{N} + \frac{3 \, \sqrt{5}}{4 \, N^3} & 
4 + \frac{5}{N^2} + \frac{5}{2 \, N^4} \end{array} \right) + \ldots \, .
\eeq
The root of the tree matrix is uniquely determined if we choose it real and symmetric. Since it is not orthogonal the transformation can change the scalar product of the eigenvectors; they become orthogonal.

The structure of the $N$ dependence is not blurred by the transformation, although the one-loop matrix now has an infinite $N$ expansion. The closest to what we have done in Section \eqref{QFT} is to look for an eigenvector of this new mixing matrix of the form $\{1, \, b\}$, i.e. we complete the single-trace state with admixtures without rescaling it. Our old mixing coefficients in \eqref{eigen73} are obviously related by the transformation \eqref{upRoot} \emph{up to an $N$ dependent rescaling} putting the first component of the eigenvector to 1. With this definition
\beq
b_1 \rar -3 \, \sqrt{5} \, , \qquad \hat b_1 \rar \frac{423 \, \sqrt{5}}{4} 
\eeq
and finally:
\beq
\braket{ (7,3)^+_c * (7,3)^+_c } \, = \, N^7 (1 + 45 \, N^{-2} - \frac{6345}{2} \, N^{-4} + \ldots)
\eeq

\subsubsection*{Degenerate pair at (8,3)}

The one-loop mixing matrix for the parity odd $L \, = \, 8, \, n \, = \, 3$ operators is
\beq
M \, = \, \left( \begin{array}{ccc} 4 & 0 & 0 \\ 0 & 6 & \frac{6}{N} \\ \frac{-2}{N} & \frac{6}{N} & 6 \end{array} \right)
\eeq
in the basis $\{ (8,3)^-, \, (8,3)^e, \, (6,3)^e (2,0) \}$. This does have a left-eigenvector $(1,0,0)$ as stated in Section \ref{QFT}. The similarity transformation by the root of the tree level matrix results to:
\beq
M \rar \left( \begin{array}{lll} 4 + \frac{1}{2 \, N^2} + \frac{209}{32 \, N^4} & \frac{5}{4 \, N^2} + \frac{4}{N^4} & \frac{-1}{N} + \frac{21}{8 \, N^3} \\[1 mm]  \frac{5}{4 \, N^2} + \frac{4}{N^4} & 6 - \frac{25}{32 \, N^4} & \frac{6}{N} + \frac{11}{8 \, N^3} \\[1 mm]  \frac{-1}{N} + \frac{21}{8 \, N^3} & \frac{6}{N} + \frac{11}{8 \, N^3} & 6 - \frac{1}{2 \, N^2} - \frac{23}{4 \, N^4} \end{array} \right) + \ldots 
\eeq
Inspecting the latter matrix one might worry whether it is still possible to complete the $(8,3)^-$ operator to a large $N$ eigenstate. In fact, there is no problem:
\beq
(8,3)^-_c \, = \, (8,3)^- - \frac{17}{8 \, N^2} \, (8,3)^e + \left( \frac{1}{2 \, N} + \frac{83}{16 \, N^3} \right) \, (6,3)^e (2,0) + \ldots  \, .
\eeq
It is clear that the state must survive because it can also be derived it by acting on the vector of coefficients $(1,0,0)$ in the other basis by multiplying with $\sqrt{S}$ and rescaling the first component. The two-point function becomes
\beq
\braket{ (8,3)^-_c * (8,3)^-_c } \, = \, N^8 \left( 1 + \frac{1}{4 \, N^2} + \frac{621}{64 \, N^4} +\ldots \right) \, .
\eeq

\subsubsection*{Degenerate pairs at (9,3)}

Let us order the complete basis as
\beq
\{ \cO^{1,2,3}, \, \cO^{1,2,4}, \, \cO^{1,2,5}, \, \cO^{1,2,6}, \, \cO^{1,2,7}, \, \cO^{1,2,8}, \, \cO^{1,3,5}, \, \cO^{1,3,6}, \, \cO^{1,3,7}, \, \cO^{1,4,7} \} \, .
\eeq
We have the paired states
\begin{eqnarray}
(9,3)_1^j & = & ( 0.019698, -0.061589\mp0.037646 \, i, -0.032810\pm0.12248 \, i, 0.14940,-0.032810\mp0.12248 \, i, \nonumber \\
&& -0.061589\pm0.037646\,  i,  0.33666, -0.28492-0.48330 \, i, -0.28492\pm0.48330 \, i, 0.25288) \, , \nonumber \\
(9,3)_2^j & = & (-0.13995, 0.24631\pm0.43613 \, i, -0.29736\mp0.22678 \, i, 0.38201, -0.29736\pm0.22678 \, i, \\ 
&& 0.24631\mp0.43613 \, i, -0.18337, 0.0070425\mp0.091441 \, i, 0.0070425\pm0.091441 \, i, 0.029342) \, , \nonumber \\
(9,3)_3^j & = & (0.38917,-0.23868\mp0.24160 \, i,-0.17869\mp0.42845 \, i,0.056392,-0.17869\pm0.42845 \, i, \nonumber \\
&& -0.23868\pm0.24160 \, i, -0.15640,0.20049\mp0.08976 \, i,0.20049\pm0.08976 \, i,0.14458) \, . \nonumber
\end{eqnarray}
The parity odd states \eqref{odd93} have the two-point functions
\begin{align}
\braket{(9,3)_1^- * (9,3)_1^-}&=\,N^9 \left(1 - 7.1317\, N^{-2} + 14.706 \, N^{-4}+\cO(N^{-6}) \right), \nonumber \\
\braket{(9,3)_2^- * (9,3)_2^-}&=\, N^9 \left(1 + 0.3985 \, N^{-2}-87.730 \, N^{-4}+\cO(N^{-6}) \right), \nonumber \\
\braket{(9,3)_3^- * (9,3)_3^-}&=\, N^9 \left(1 - 33.267 \, N^{-2}+414.02 \, N^{-4}+\cO(N^{-6}) \right) \, , \\
\braket{ (9,3)_1^- * (9,3)_2^- } & = \, N^9  \left( 0 + 5.9542 \, N^{-2} - 29.236 \, N^{-4} + O(N^{-6}) \right) \, , \nonumber \\
\braket{ (9,3)_1^- * (9,3)_3^- } & = \, N^9 \left( 0 - 0.7722 \, N^{-2} - 0.4182 \, N^{-4} + O(N^{-6}) \right) \, , \nonumber \\
\braket{ (9,3)_2^- * (9,3)_3^- } & = \, N^9 \left( 0 -1.4489 \, N^{-2} + 50.853 \, N^{-4} + O(N^{-6}) \right) \, . \nonumber
\end{align}
Further,
\begin{eqnarray}
\braket{ (9,3)_1^- * (7,3)^- (2,0) } & = & \phantom{-}0.9411 \, N^8 - 13.515 \, N^6 + O(N^4) \, , \nonumber \\
\braket{ (9,3)_2^- * (7,3)^- (2,0) } & = &-5.1929 \, N^8 + 79.646 \, N^6 + O(N^4) \, , \\
\braket{ (9,3)_3^- * (7,3)^- (2,0) } & = &-2.4795 \, N^8 + 51.496 \, N^6 + O(N^4) \nonumber
\end{eqnarray}
and
\begin{eqnarray}
\braket{ (9,3)_1^- * (6,3)^e (3,0) } & = & \phantom{-}2.9140 \, N^8 -  12.874 \, N^6 + O(N^4) \, , \nonumber \\
\braket{ (9,3)_2^- * (6,3)^e (3,0) } & = & \phantom{-}5.7868 \, N^8 -  60.161 \, N^6 + O(N^4) \, , \\
\braket{ (9,3)_3^- * (6,3)^e (3,0) } & = &-0.1467 \, N^8 - 11.267 \, N^6 + O(N^4) \, . \nonumber
\end{eqnarray}
In combination with formulae \eqref{sq6330}, \eqref{sq7320}, \eqref{mix73206330} of Appendix B these yield the tree matrix $S_{ij}$ whose root defines the similarity transform to orthogonal states. The two-point functions for the $(9,3)^-_{j,c}$ states are
\begin{align}
\braket{ (9,3)_{1,c}^- * (9,3)_{1,c}^-} &= N^9\left(1  + 5.6881 \, N^{-2}+ 7.5170 \, N^{-4} + O(N^{-6}) \right) \, , \nonumber \\ 
\braket{ (9,3)_{2,c}^- * (9,3)_{2,c}^- } &= N^9\left(1  + 174.91 \, N^{-2}- \phantom{.}10337 \, N^{-4} + O(N^{-6}) \right) \, , \\ 
\braket{ (9,3)_{3,c}^- * (9,3)_{3,c}^- } &= N^9\left(1  + 0.3980 \, N^{-2}+ 13.221 \, N^{-4} + O(N^{-6}) \right) \, . \nonumber
\end{align}

\section*{Appendix B: disconnected double-trace two-point functions}

In the disconnected part of the two-point functions of $(6,3)^e (3,0)$ we encounter the norm
\beq
\braket{ (6,3)^e * (6,3)^e } \, = \, 1 * (N^6 - 6 \, N^4 + \ldots) + 1 * (N^4 + \ldots) \, = \, N^6 - 5 \, N^4 + \ldots
\eeq
where we have separated by sphere and torus diagrams. The torus part involves graphs that can be marked on Figure 3, Panel 1 with two length 6 and two identity operators as in \cite{colourDressed}. The relevant sets of edge widths are
\beq
\{0, 1, 1, 4\}, \, \{0, 2, 2, 2\}, \, \{1, 1, 1, 3\}, \, \{1, 1, 2, 2\}, \, \{1, 2, 1, 2\} \, .
\eeq
In field theory as well as in integrability, the amplitudes for the first and third of these vanish.
The hexagon computation has to be scaled up by a factor $\sqrt{L_1 \, L_2} \, = \, 6$ upon which the second and the fifth case have to receive extra factors $1/3, \, 1/2$, respectively, due to their three- and twofold cyclic symmetry, cf. \cite{colourDressed}.

Trivially,
\beq
\braket{ (3,0) * (3,0) } \, = \, 1 * \frac{(N^2-1)(N^2-2)}{N} + 1 * \frac{2 \, (1-N^2)}{N} \, = \, N^3 - 5 \, N + \ldots
\eeq
so that the product yields $N^9 - 10 \, N^7 + \ldots$ as stated in the paragraphs after \eqref{sampleRibbon} for the $\{6,0,3,0\}$ double-trace graph. From the other formulae in Section \ref{tessel} it follows 
\beq
\braket{ (6,3)^e (3,0) * (6,3)^e (3,0) } \, = \, N^9 + 2 \, N^7 + \ldots \label{sq6330}
\eeq

As before, we will compute the $\{7,0,2,0\}$ sphere and torus parts of the $(7,3)^- (2,0)$ two-point functions from the overlaps of the two Bethe states $(7,3)^j$. The sphere part is obviously diagonal and comes with the colour factor $(N^7 - 7 \, N^5 + \ldots)(N^2 - 1) \, = \, N^9 - 8 \, N^7 + \ldots$. In the torus part of $\braket{ (7,3)^j * (7,3)^k }$ we find the colour ribbon graphs
\beq
\{0, 1, 1, 5\}, \, \{0, 2, 2, 3\}, \, \{1, 1, 1, 4\}, \, \{1, 1, 2, 3\}, \, \{1, 1, 3, 2\}, \, \{1,
2, 1, 3\}, \, \{1, 2, 2, 2\}
\eeq
which evaluate to:
\beq
N^5 \, \{ -1,1,1,1,1,1,1\} 
\eeq
Now, w.r.t. this basis:
\begin{eqnarray}
\braket{ (7,3)^1 * (7,3)^1 } = & \{ 2, \, 0, \, 2, \, -1 - \sqrt{15} \, i, \, -1 + \sqrt{15} \, i, \, 2, \, -1\} &\rightarrow - N^5 + \ldots \nonumber \\
\braket{ (7,3)^2 * (7,3)^2 } = & \{ 2, \, 0, \, 2, \, -1 + \sqrt{15} \, i, \, -1 - \sqrt{15} \, \, i, \, 2, \, -1\} &\rightarrow - N^5 + \ldots \\
\braket{ (7,3)^1 * (7,3)^2 } \, = & \braket{ (7,3)^2 * (7,3)^1 } \, = \ \{0,\,  2,\,  0, \, 3, \, 3, \, 0, \, -2\} &\rightarrow \, 6 \, N^5 + \ldots \nonumber
\end{eqnarray}
Again, the integrability results for the torus need to be scaled up by $\sqrt{L_1 L_2} \, = \, 7$ to find a match, though there are no extra factors due to cyclic invariance here. For the first Bethe state we must insert an extra sign. In conclusion, the diagonal leading $N$ disconnected part of equation \eqref{mixing73} receives a correction
\beq
- 3 \, N^7 \left( \begin{array}{rr} 3 & -2 \\ -2 & 3 \end{array} \right) \, .
\eeq
Collecting terms:
\beq
\braket{ (7,3)^- (2,0) * (7,3)^- (2,0) } \, = \, N^9 - 7 N^7 + \ldots \label{sq7320}
\eeq
For completeness, we recall:
\beq
\braket{ (6,3)^e (3,0) * (7,3)^- (2,0) } \, = \, - 2 \, \sqrt{6} \, N^7 + \ldots \label{mix73206330}
\eeq

Third, for the $(7,3)^+, \, \cO_\perp$ mixing example we need the norm of the (5,3) state to torus order. We will use the torus diagram Figure 3, Panel 1 with identity insertions at points 3,4 with transverse magnons. There are only three contributing tree diagrams,
\beq
\{0,0,0,5\}, \, \{0,1,1,3\}, \, \{1,1,1,2\} \, , \label{tuples53}
\eeq
where we include the sphere contribution as an additional test although it is already covered by \eqref{norm53}. 

This offers an opportunity to apply the rules for twist developed above. We should not put twist on the edges connecting the two length five operators by what was said before. On the other hand, hiding the twist on the edges of width zero the $n$ coefficients tend to drop. Hence, starting both partitions in the centre of the figure so that edge A is crossed first it should be possible to compute without any regulator. And indeed, putting $a_1 \, = \, 0, \, a_2 \, = \, 1$ and normalising as in the derivation of \eqref{norm53} we obtain
\beq
- \frac{(a_3-a_4)}{a_3 \, (1-a_4)} \ \{1, 1/5, \, 1/5 \} \ .
\eeq
The complete space time factor would likely be $-a_{34}/(a_{12}^5 \, a_{13} \, a_{24})$. It comes as a surprise because of the wrong sign, but also because the positions of the identity operators do not drop, and because it has a preferred association between points 1,3 and 2,4, respectively, which might arise from the turning sense of the partitioning\footnote{A non-trivial space time factor arises also from the torus diagram in Figure 3, Panel 3. However, in that instance the dependence on the fictious operators' positions drops in the two-point limit.}. As the computation impeccably yields the correct amplitudes (as usual, the true torus part needs an extra factor $\sqrt{L_1 L_2} \, = \, 5$) we clearly need to eliminate the space time factor and the additional sign. The point deserves future attention.

The three colour factors associated with \eqref{tuples53} are $N^5 - 5 \, N^3, \, -N^3, \, N^3$ up to the relevant order, so that the true torus part does not even come in. Together with the
$\braket{(2,0) * (2,0)}$ part $N^2-1$ the first entry in the and the table after \eqref{beforeTheTable} should thus pick up a colour factor $N^7 - 6 \, N^5 + \ldots$

\section*{Appendix C: the correlators of Section \ref{descCor}  with longitudinal magnons} 

We repeat the computations identically, though with $Y$ magnons. For ease of comparison we stick to the partitions yielding the tables in Section \ref{descCor}. As we know by now, different choices would only result in reparametrisations, 

\subsection*{$\mathbf{\langle(5,3)(3,0)*(5,3)(3,0)\rangle}$ with longitudinal magnons}

For the \emph{empty square} the $c_0$ independence condition \eqref{cIndep} yields
\beq
\na \, = \, \naa.
\eeq
as before. With that
\vskip 0.2 cm
\begin{center}
\begin{tabular}{l|c|l} 
$\la\lb\lc\ld$ & QFT & hexagon tiling \\
\hline
5020 & 0 & $\phantom{10 *} \ \ \frac{1}{4} \, (1 - \na)(1 + \na)(4 + 5 \, \na) (\na + \nd)$ \\[1mm]
4111 & -2 & $10 * \frac{1}{20} (1 + \na) (2 - 5 \, \na) (4 + 5 \, \na) (\na + \nd)$ \\[1mm]
3202 & -2 & $10 * \frac{1}{20} (2 + 5 \, \na)(1 - 8 \, \na - 5 \, \na^2)(\na + \nd)$ \\
\end{tabular}
\end{center}
A ratio of the second and third entry shows that there is no universal, real solution for $\na, \, \nd$. The most likely interpretation will then be to put $\na \, = \, 0$ and to have non-universal $\nd \, = \, \{0, \, 1/2, \,  2 \}$. The twist trick unfortunately loses all predictive power.

With identical $n$ coefficients at both ends of the equator lines, the \emph{belt} diagram does not have any constraint from $c_0$ independence. Yet, we do see the $\na, \, n_{C,3}$ parameters now. The existence of the point identification limit imposes a new constraint:
\beq
\na \, = \, 0 \qquad \lor \qquad n_{C,3} \, = \, 0
\eeq
These yield exchangeable results, so let us assume $n_{C,3} \, = \, 0$. We find
\vskip 0.2 cm
\begin{center}
\begin{tabular}{l|c|l} 
$\{\lh,\lv\}\la\lb\lc\ld$ & QFT & hexagon tiling \\
\hline
\{3,0\}0202 & -2 & $10 * \frac{1}{10} (1 + 9 \, \nh + 10 \, \nh^2) (1 - \na) $ \\[1mm]
\{2,1\}0202 & 4 & $10 * \frac{1}{10} (1 - 3 \, \nh + 10 \, \nh^2) (1 - \na)$ \\
\end{tabular}
\end{center}
There is no universal solution here either. One would presumably put $\nh \, = \, 0$ and tune $\na$.

\subsection*{$\mathbf{\langle(5,3)(2,0)*(5,2)(2,1)\rangle}$ with longitudinal magnons}

The \emph{empty square} has the $c_0$ independence constraint $\nd \, = \, 2 \, (1 - \nc)$ as before. We compute:
\begin{center}
\begin{tabular}{l|c|l} 
$\la\lb\lc\ld$ & QFT & hexagon tiling \\
\hline
5020 & $-\sqrt{2}$ & $\phantom{10 *} \ \, \frac{\sqrt{2}}{4} (1 + \na) (4 + 5 \, \na) (-1 + \nc)(2 + \na - 2 \, \nc)$ \\[1mm]
4111 & $-5 \sqrt{2}$ & $10 * \frac{\sqrt{2}}{40} (4 + 5 \, \na) (-5 - 5 \, \na + 6 \, \nc + 10 \, \na \nc)(2 + \na - 2 \, \nc)$ \\[1mm]
3202 & $ - 2 \sqrt{2}$ & $10 * \frac{\sqrt{2}}{20} (2 + 5 \, \na) (-2 + 3 \, \nc + 5 \, \na \nc) (2 + \na - 2 \, \nc)$ \\
\end{tabular}
\end{center}
In the \emph{belt} case we fall upon the known $c_0$ independence constraint $\nh \, = \, -\frac{1}{2} - \frac{\na}{2} + \nc$. The point identification limit $2 \rar 1, \, 4 \rar 3$ requires $\na \, \nc \, (11 + 5 \, \na - 10 \, \nc)\, = \, 0$ for the $\{3,0\}0202$ graph and $\na \, \nc \, (3 + 5 \, \na - 10 \, \nc)$ for $\{2,1\}0202$. So for both graphs we have three possibilities:
\begin{center}
\begin{tabular}{l|c|l} 
$\{\lh,\lv\}\la\lb\lc\ld$ & QFT & hexagon tiling \\
\hline
\{3,0\}0202 & $-2 \sqrt{2}$ & $10 * \frac{\sqrt{2}}{20} (1 + 5 \, \na + \nc + 5 \, \na \nc - 10 \, \nc^2) (1 - 2 \, \nc)|_{\na \, = \, 0}$ \\[1mm]
& & $10 * \frac{\sqrt{2}}{20} (1 + 5 \, \na + \nc + 5 \, \na \nc - 10 \, \nc^2) (1 - \na)|_{\nc \, = \, 0}$ \\[1mm]
& & $10 * \frac{\sqrt{2}}{10} (6 + 5 \, \na)(1 - \na)$ \\[1 mm]
\{2,1\}0202 & 0 & $10 * \frac{\sqrt{2}}{20} (-5 - 5 \, \na + 13 \, \nc + 5 \, \na \nc - 10 \, \nc^2) (1 - 2 \, \nc) |_{\na \, = \, 0}$ \\[1mm]
& & $10 * \frac{\sqrt{2}}{20} (-5 - 5 \, \na + 13 \, \nc + 5 \, \na \nc - 10 \, \nc^2) (1 - \na) |_{\nc \, = \, 0}$ \\[1mm]
& & $10 * \frac{\sqrt{2}}{50} (-2 + 5 \, \na) (1 - \na)$ \\
\end{tabular}
\end{center}
With the exception of the $\{2,1\}$ case in the second table none of this can be satisfied with  likely values for $\na$ or $\nc$.

\subsection*{$\mathbf{\langle(5,2)(2,1)*(5,2)(2,1)\rangle}$ with longitudinal magnons}

We have identified $n_{C,3} \, = \, \nc$ whereby no constraint arises from \eqref{cIndep}.
\vskip 0.2 cm
\begin{center}
\begin{tabular}{l|c|l} 
$\la\lb\lc\ld$ & QFT & hexagon tiling \\
\hline
5020 & -1 & $\phantom{10 *} \ \, - (1 - \nc) (1 - 2 \, \nb - 2 \, \nc) $ \\[1mm]
4111 & -7 & $10 * \frac{-1}{10} (5 - 6 \, \nc) (1 - 2 \, \nb - 2 \, \nc) $ \\[1mm]
3202 & -4 & $10 * \frac{-1}{10} (2 - 3 \, \nc) (1 - 2 \, \nb - 2 \, \nc) $ \\
\end{tabular}
\end{center}
Forming ratios we find $\nc \, = \, 2$. From the absolute normalisation it then follows that $\nb \, = \, -1$. For once there is a fairly appealing solution!

The \emph{belt around the belly} comes with the constraint $\naa \, \nc \, = \, 0$ if we let the magnons of the two vacuum descendents cross the edges $A, \, C$, respectively, on their path through the entangled state. Assuming $\nc \, = \, 0$ the amplitudes are:
\vskip 0.2 cm
\begin{center}
\begin{tabular}{l|c|l} 
$\{\lh,\lv\}\la\lb\lc\ld$ & QFT & hexagon tiling \\
\hline
\{3,0\}0202 & -4 & $10 * \frac{1}{5} (1 - 2 \, \naa)$ \\[1mm]
\{2,1\}0202 & +4 & 0 \\
\end{tabular}
\end{center}
The $\{3,0\}$ case is trivial to satisfy --- and the coefficient $3/2$ perhaps still credible --- but the second row is a blatant contradiction.

\section*{Appendix D: a pretty hexagon amplitude for longitudinal magnons}

All $Y$ hexagon amplitudes with magnons on only two of the physical edges directly evaluate to products of $A$ and $(A-B)/2$ elements of the $PSU(2|2)$ scattering matrix \cite{psu22}. Putting one of the excitations onto the third egde radically changes the picture: the scattering involves all 10 elements of the $S$ matrix and yields a large sum of terms. The most complicated amplitude of the type we encounter in the present context is
\beq
\cA_{3|1|2} \, = \, \langle {\mathfrak h} | Y_1 Y_2 Y_3 | Y_4 | Y_5 Y_6 \rangle \, = \, -
\langle {\mathfrak h} | Y_1^{4 \gamma} Y_2^{4 \gamma} Y_3^{4 \gamma} \, \bar Y_4^{2 \gamma} \, Y_5 \, Y_6 \rangle
\eeq
At tree level, it turns out to have a concise decomposition over the particle creation poles:
\begin{eqnarray}
\cA_{3|1|2} & = & i \frac{S_{21} S_{31} }{u_{14}} \, \frac{h_{32} h_{65} }{h_{25} h_{26} h_{35} h_{36}} +  i \frac{S_{32}}{u_{24}} \, \frac{h_{31} h_{65}}{h_{15} h_{16} h_{35} h_{36}} + i \frac{1}{u_{34}} \, \frac{h_{21} h_{65}}{h_{15} h_{16} h_{25} h_{26}} + \label{decomp} \\
& & i \frac{1}{u_{54}} \, \frac{h_{21} h_{31} h_{32}}{h_{16} h_{26} h_{36}} - i \frac{S_{65}}{u_{64}} \, \frac{h_{21} h_{31} h_{32} }{h_{15} h_{25} h_{35}} + \frac{h_{21} h_{31} h_{32} h_{65}}{h_{15} h_{16} h_{25} h_{26} h_{35} h_{36}} \nonumber
\end{eqnarray}
where $u_{ij}$ is a rapidity difference and the tree level $SU(2)$ S matrix is defined in \eqref{SU2mat}. Finally, the tree level dressing factor is \cite{BKV}
\beq
h_{ij} \, = \, \frac{u_i - u_j}{u_i - u_j - i} \, .
\eeq
Note that formula \eqref{decomp} singles out the poles in $u_4$ and the finite part behind; it is somewhat reminiscent of \cite{bcfw} in the amplitude literature. 

For a proof one might start refining the triangulation by inserting an identity operator in the middle of the hexagon. We obtain a partition into 64 terms with splitting factors consisting of $S$ matrices only because the new edges are of width zero. Every new hexagon amplitude has excitations at only two physical edges and thereby factors into products of $h$ factors as in \eqref{decomp}. The argument is not sufficient to explain the still simpler final form of the amplitude, though.

For longitudinal cases that do not have one preferred magnon (here the one that is alone on its edge) we have so far not been able to spot a similar pattern. Such techniques would be particularly useful in manipulating non-vanishing amplitudes for transverse magnons, which are typically bulky sums.

\end{document}